\documentclass[a4paper,11pt]{article}
\pdfoutput=1 

\usepackage{jheppub} 

\usepackage[T1]{fontenc} 

\usepackage{mathtools}
\usepackage{tikz} 
\usepackage{tikz-3dplot}
\usetikzlibrary{positioning}
\usetikzlibrary{arrows}
\usetikzlibrary{decorations.pathreplacing}
\usetikzlibrary{shapes.geometric}
\usepackage{mathtools}
\def\W{\mathcal{W}}

\def\a{\alpha}
\def\b{\beta}
\def\F{\textbf{F}}

\def\F{\textbf{F}}
\def\AF{\textbf{AF}}
\def\inf{\infty}
\def\bsp{\begin{split}}
\def\esp{\begin{split}} 
\def\N{\mathcal{N}}
\def\Z{\mathbb{Z}}

\def\tk{\tilde{k}}

\def\N{\mathcal{N}}

\def \redcirc{{\color{red}{\bigcirc}}}
\def\bluecirc{{\color{blue}{\bigcirc}}}

\def\tL{\widetilde{L}}

\def\PE{\text{PE}}

\def\tx{\tilde{x}}
\def\q{ \mathfrak{q}}

\def\G{{\mathcal{G}}}
\def\C{{\mathcal{C}}}
\def\redLI{{\color{red}L_1}}
\def\redLII{{\color{red}L_2}}
\def\gauge{
\tikz[baseline={(0,0)}]{ 
    \fill (0,0) circle (3pt);
} }

\tikzset{gauge/.style={circle,fill,inner sep=2pt}}
\tikzset{matter/.style={fill=gray, inner sep=2.4pt,regular polygon,regular polygon sides=4}}

\def\doubsquare{   
\tikz[baseline={(0,-0.1)}]{ 
\draw[thick] (-0.16,-0.16) rectangle (0.16,0.16)  ;
\draw[thick] (-0.1,-0.1) rectangle (0.1,0.1)  ;
} }
\def\dbox{   
\tikz[baseline={(0,-0.1)}]{ 
\draw[thick,fill=white] (-0.14,-0.14) rectangle (0.14,0.14)  ;
\draw[thick] (-0.09,-0.09) rectangle (0.09,0.09)  ;
} }
\def\sqbox{   
\tikz[baseline={(0,-0.1)}]{ 
\draw[thick,fill=white] (-0.12,-0.12) rectangle (0.12,0.12)  ;
} }

\usepackage[english]{babel}

\title{\boldmath 
Fusions and Dualities for 3d Theories $T[M_3]$
}

\author[1,2]{Shi Cheng}

\affiliation[1]{
Shanghai Institute for Mathematics and Interdisciplinary Sciences (SIMIS), 200433 Shanghai, China}
\affiliation[2]{
Department of Physics and Center for Field Theory and Particle Physics, Fudan University, 20005, Songhu Road, 200438 Shanghai, China}

\emailAdd{mirror2718@gmail.com}

\abstract{

\vspace{4mm}
\noindent

We study 3d theories determined by three-manifolds. Previously, we found that some basic 3d dualities relate to the surgeries of three-manifolds and defined gauge circles and matter circles. In this note, we discuss some operations including handle slides, gaugings and flips of mass parameters, and the corresponding geometric interpretations for these operations. 
We note that a fusion identity could describe the fusion, or in other words, the connected sum of matter circles. Many abelian dualities could arise from this fusion and inherit geometric interpretations thereof.

}

\begin{document} 
\maketitle
\flushbottom

\section{Introduction}
\label{sec:intro}

We study the geometric realizations of 3d gauge theories arise from  the DGG correspondence \cite{Dimofte:2011ju} and \cite{Terashima:2011qi,Cecotti:2011iy}. Although there are already many distinguishing works, we have not reach the understanding of how to geometrically engineer 3d theories through compact three-manifolds. In \cite{Gadde:2013aa}, compact three-manifolds are used for the compactification of M5-branes to get 3d theories $T[M_3]$, while  it is unknown that what the theories $T[M_3]$ are, as these theories are believed to be non-lagrangian. Fortunately, it is clear that if one puts a single M5-brane on the three-manifold, what one gets is an abelian 3d $\N=2$ theory without coupled chiral multiplets. It looks a trivial theory, however the further study shows even abelian theories are highly non-trivial, partially because of the existence of the duality group $SL(2,\Z)$ found in \cite{Witten:2003ya}. In the presence of chiral multiplets, a subset of this duality group still preserves as is shown in \cite{Dimofte:2011ju,Cheng:2023ocj}.  As for the geometric construction of abelian theories with chiral multiplets, the compact three-manifolds should be decorated by putting Ooguri-Vafa (OV) defects in the cotangent bundles $T^*M_3$ to introduce flavor symmetries for the chiral multiplets \cite{Ooguri:1999bv,Aganagic:2012hs,Cheng:2023zai}. One should also put M5-branes on this OV defects to engineer flavor symmetries, and then the chiral multiplets are realized by the M2-branes stretching between the OV defects and the base three-manifold $M_3$ that gives gauge symmetries. This configuration is checked through M-theory/IIB string duality for Lens spaces, and we believe it is true for at least abelian theories on any compact three-manifolds given by surgeries along links. 

We also find some 3d dualities can be interpreted as some extended Kirby moves in the presence of OV-defects \cite{Cheng:2023zai}. These OV defects give Wilson loops if considering the dual 3d Chern-Simons theories through 3d-3d correspondence; see e.g.\cite{Aganagic:2012hs}. We defined some notations for descriptions.
The intersections between OV-defects and the three-manifolds are some unknots which we call matter circles as they encode chiral multiplets through M2-branes. The gauge symmetries come from the surgeries along links of unknots, which we call gauge circles. With these geometric objects in mind, one can
discuss the the geometric transformations of three-manifolds in the presence of OV defects. The original Kirby moves are found in \cite{kirbymove} that there are two kinds of equivalent surgeries, including blow-up/down and handle slides, which are also called the first type of Kirby moves and the second type of Kirby moves respectively. In the previous work \cite{Cheng:2023zai}, we have not clearly discussed handle slides. In this note, we will discuss the handle slides in the presence of matter circles. In this note we find that it is possible to perform handle slides on both gauge circles and matter circles. The first case could change locations  of matter circles. The second case is still unclear to us, but a special case of it is the connected sum of matter circles, which is already quite mysterious and powerful, so we call it fusion. The geometric interpretation of the fusion is shown through the figure in \eqref{fusioncirc}. From this fusion, we derive all 3d abelian dualities that we know, and these dualities are confirmed by the equivalence of vortex partition functions which are written as identities involving $\q$-Pochhammer products. We use some physical operations to find descendent dualities from the fusion identity \eqref{fusion}, including gauging mass parameters (flavor symmetries) which is the key in understanding vortex partition functions. Other operations include taking different mass and flux number limits, flipping signs of mass parameters as well as orientations of gauge circles and matter circles. These operations are geometrically interpreted as adding surgery circles, taking special lengths for surgery circles, and changing orientations respectively.
We find the SQED-XYZ duality arises immediately from the fusion identity, and then S-duality and ST-duality arise from SQED-XYZ duality thereof. These descendent dualities form a complicated web \eqref{fusiontoST} which even encodes a hiding algebraic structure.

From the fusion identity, as we have just mentioned that the cubic superpotential in the mirror dual pair SQED-XYZ arises if taking a large mass limit, which indicates that the process of the fusion of matter circles encodes the cubic superpotential. The fusion should be a closed geometric interpretation for cubic superpotentials.  One can do infinite many fusions to a three-manifold and correspondingly add infinite many cubic terms to the superpotential.
We need to mention that in this process handle slides of gauge circles simplify the analysis. We summarize the relation between fusion and three types of superpotential triangles as follows:
\begin{align}
\text{fusion of matters \eqref{fusioncirc}}~~\longrightarrow~~ \text{cubic superpotential \eqref{unlinking},\eqref{linkingform},\eqref{exoticcase}} \,.
\end{align}

The treatment of the vortex/quiver partition functions are also meaningful in the perspective of mathematics. Firstly, the 3d duality webs \eqref{fusiontoST}  given by fusion identities provide  relations between many $\q$-series.
The fusion identity is the origin of many identities for $\q$-series, including the most familiar ones and these for describing physical dualities. The network between the $\q$-series that we known are shown in \eqref{fusiontoW} and \eqref{fusiontoST}, and there could be more. In addition, this note is a computation support to various results in \cite{Cheng:2023zai}.

The outline of this note is as follows.  In section \ref{quiver}, we show the quiver generating function and its extended form as the vortex partition function. In section \ref{sec:gaugemass}, we discuss how to gauge flavor symmetries by shifting mass parameters, which is the main technique that makes possible the computation in this note. In section \ref{sec:flips}, we discuss a delicate flip of mass parameters that involves the orientations and twists of  three-manifolds with $\q$-Pochhammer products. In section \ref{sec:STmove}, we analyze the ST-duality by using the techniques we find, which require us to properly reorganize the vortex (quiver) partition function to math with the corresponding 3d dualities. In section \ref{sec:handleslides}, we discuss the influence of handle slides of surgery circles on matter circles, which we find could change the charges and locations of matter circles but are still equivalent geometric transitions. 
In section \ref{sec:fusion}, we notice that the fusion identity could describe the connected sum of matter circles, which has not been acknowledged in literature. This fusion derives many other dualities including the S-duality, ST-duality and their gauged versions, and can be interpreted as a particular Hanany-Witten transition in terms of brane webs. In section \ref{sec:triangles}, we discuss the influence of handle slides on superpotential triangles that found in \cite{Ekholm:2019lmb,Cheng:2023ocj}, and noticed that the fusion could lead to these superpotential triangles. This indirectly shows that the cubic superpotential arises from fusion or in other words the connected sum of matter circles. In section \ref{sec:structures}, we draw a duality web to discuss the relations between different dualities that come from fusion identity. In section \ref{sec:Witten}, we try to connect fusion to 3d topological field theroy description in Witten's work \cite{witten89}, which shows a subtle difference. In section \ref{sec:open}, we list some unsolved problems.

\vspace{4mm}
\textit{
As this paper was being finalized, we  became aware of the section \ref{sec:STmove} and section \ref{sec:triangles} of this note have overlapped results with the section 2 and section 3 of \cite{Gaiotto:2024ioj}.  }

\section{Surgery and matter for 3d theories} \label{quiver}

\subsection{Quiver/vortex generating functions}
Quivers in this note are defined as symmetric matrices that encoding a particular generating function \cite{Kontsevich:2010px}. Quivers describe the BPS states in a different context, but one should be careful if trying to describe these as the BPS states in 3d gauge theories or open topological strings.

The quiver is denoted by the matrix $C_{ij}$, from which the quiver generating function can be defined and take the form 
\begin{equation}
P(C_{ij}; x) = \sum_{n_1, n_2 , \cdots, n_N=0}^{\infty} ( - \sqrt{\q})^{ \sum_{i,j=1}^{N} C_{ij} n_i n_j  }  \prod_{i=1}^{N} \frac{ x_i^{n_i} }{  (\q, \q)_{n_i} } \,.
\end{equation}
The quiver can be physically interpreted as the effective mixed Chern-Simons level for 3d $\N=2$ abelian gauge theories if comparing quiver generating functions with vortex partition functions.
We notice that in order to precisely match with 3d gauge theories, one-loop contributions should be added and generic charges of chiral multiplets can be turn on. Then we propose an extended quiver generating function:
\begin{align}\label{genericquivers}
Z_{K_{ij}}(Q_\rho, x_i):=& 
\prod_{\rho=1}^{N_f} (Q_\rho, \q)_\inf   \cdot
 \sum_{n_1, \cdots, n_{N_c=0}}^{\infty} 
 ( - \sqrt{\q})^{ \sum_{i,j=1}^{N} K_{ij} n_i n_j  }  \frac{ \prod_{i=1}^{N_c}  x_i^{n_i} }{  \prod_{\rho=1}^{N_f} 
 (Q_\rho, \q)_{\sum_{i=1}^{N_c} q_i^\rho n_i^\rho  } }  \nonumber \\ 
 =& \sum_{n_1, \cdots, n_{N_c=0}}^{\infty} 
 ( - \sqrt{\q})^{ \sum_{i,j=1}^{N} K_{ij} n_i n_j  }  { \prod_{i=1}^{N_c}  x_i^{n_i} }{  \prod_{\rho=1}^{N_f} 
 \Big(Q_\rho^{\sum_{i=1}^{N_c} q_i^\rho n_i^\rho  }, \q \Big)_\inf }
 \,,
\end{align}
which is just the form for vortex partition functions of 3d abelian theories. The $q_i^\rho$ is the charge of the $\rho$-th matter under $i$-th gauge group $U(1)$, and $Q_\rho$ are the K\"ahler parameters that correspond to the mass parameters of the $\rho$-th matter through the relation $Q \sim \exp(\text{vol} (\C)) = \exp( m)$, where $\C$ is the volume of the M2-branes engineer the vortices by wrapping on the holomorphic disc $\C$ in the context of M-theory.

\vspace{2mm}\noindent
\textbf{The advantages over sphere partition functions.}
In \cite{Cheng:2023ocj}, the abelian 3d theories are studied by sphere partition functions.
The quiver generating functions as the vortex partition functions of abelian 3d theories have various advantages over sphere partition functions. Firstly, the sphere partition functions are sometimes not obvious enough for reading off physical informations, and the correspondence between parameters through dualities are not very precise if not taking residues in computation.
Vortex partition functions, however, can very well keep tack of the transformations between parameters, and one can even see if dual theories are equivalent up to some free singlets. The gauging processes that we will discuss in section \ref{sec:gaugemass}
 are also much more non-trivial in vortex partition functions than sphere partition functions. The equivalent vortex partition functions for dual theories are identities involving $\q$-Pochhammer products. Last but not the least, the vortex partition functions are more directly related to knot invariants, three-manifold invariants and geometric transformations, such as Kirby moves.
In e.g.\cite{Kucharski:2017ogk,Chung:2023qth}, some identities for Pochhammer products are used to rewrite the knot invariants in the form of quiver generating functions. In this note, we give a physical and geometric interpretation for these and many other identities.

\subsection{ Gauge circles and matter circles}
In this section, we briefly review the results that we obtained in \cite{Cheng:2023zai}, which is on the surgery description of abelian 3d $\N=2$ theories $T[M_3]$ obtained by wrapping a single M5-brane on a plumbing manifold $M_3$. 
 In addition, chiral multiplets are also included by adding Ooguri-Vafa defects that are given by wrapping other M5-branes on Lagrangian sub-manifolds of the type discussed in \cite{Ooguri:1999bv}. 
This Ooguri-Vafa defect has the topology of the solid torus $K \times \mathbb{R}^2 \subset T^*M_3$ and $K$ as a knot is the intersection between OV defect and the base three-manifold, namely $(K \times \mathbb{R}^2) \cap M_3 =K$. One can also lift this OV defect into the cotangent bundle and hence M2-branes wrapping on $K \times I$ emerge and stretch between the OV defect and $M_3$, which is basically the brane configuration of $\text{M5-M2-M5'}$. A rough figure below can illustrate,
\begin{equation}
\begin{aligned}
\begin{tikzpicture}
 \draw[thick,red] (0,0)--(1,0)  (0,0.2)--(1,0.2) ;
 \draw[thick,red] (0,0.1) ellipse (0.05 and 0.1)   (1,0.1) ellipse (0.05 and 0.1)  ;
 \node at (0.45,0.2) {\small M2} ;
  \node at (-0.4,1.1) {\small M5} ;
    \node at (1,1.1) {\small M5'} ;
       \node at (1.6,0.3) {\small OV} ;
        \node at (-0.7,0.3) {\small $M_3$} ;
 \draw[thick,blue] (0.2,-0.5) --(0.2,0.5)--(-0.3,0.8)--(-0.3,-0.2)--(0.2,-0.5) ;
  \draw[thick] (1.2,-0.5) --(1.2,0.5)--(1-0.3,0.8)--(1-0.3,-0.2)--(1.2,-0.5) ;
\end{tikzpicture}
\end{aligned}
\end{equation}
where $M_3$ is a compact three-manifold and hence gives gauge symmetries, and the OV defect is non-compact and hence gives flavor symmetries.
In this note and \cite{Cheng:2023zai}, we only know how to address the case that $K$ is an unknot $\redcirc$. 
The author found that the delicate surgery structures of $M_3$ determine the 3d theories. In short, the $M_3$ is given by surgeries along links of circles, namely $M_3 =S^3 \backslash ( L_1\cup L_2 \cup \cdots ) $. The surgery is the process that one cuts out the neighborhood of each $L_i$ and then fills in solid tori along the boundaries, and the gluing maps are elements in the mapping class group of $SL(2,\mathbb{Z})$ and are labeled by framing numbers as integers. In \cite{Gadde:2013aa}, it is argued that each surgery component $L_i$ gives a gauge group $U(1)_{k_i}$, and in \cite{Cheng:2023zai}, the OV defects are introduced to encode matters, which says that each chiral multiplet is given by the M2-brane in the above brane configuration, and the charge of the matter is the winding number between matter circle $\redcirc$ and gauge circle $L_i$ which is denoted by the blue circle $\bluecirc$ or a black node $\bullet_k$ in this note. In short, the notation is
\begin{equation}
U(1)_k:~~~~\bluecirc_{k} \,,~~~~ \bullet_k  \,,\qquad   
~~\text{chiral multiplet} :~~~~\redcirc \,,~~~~\sqbox\,.
\end{equation}

The above conclusion is obtained by comparing M-theory with 3d brane webs in IIB, and by comparing the 3d dualities with equivalent surgeries of various types on three-manifolds.
In addition, we remind that the mass parameters and FI parameters are interpreted as the lengths of matter circles and gauge circles respectively.

In this note, we continue this surgery and defect realization of 3d theories through three-manifolds. We will give geometric interpretations for gauging of flavor symmetries, flips of the mass parameters, as well as handle slides of extended version (fusion). These more delicate geometric realizations of physical operations provide a relatively complete and systematic method for analyzing abelian 3d theories.

\section{Gauging global symmetries}\label{sec:gaugemass}

In \cite{Cheng:2023ocj}, we apply gauging as a physical operation on basic dual theories to derive the gauged versions. For instance, ST-move is the gauged ST-duality that is a duality between a free chiral multiplet $\Phi$ and the $U(1)_{\pm1/2}+ 1\Phi$. Another example is the SQED-XYZ duality \cite{Intriligator:1996ex}, which involves a cubic superpotential. However, we have not clearly discussed the gauging of mass parameters, as basically it is not an equivalent operation for 3d theories.
In addition, 
there is a puzzle in particular for the gauging of mass parameters $Q$ in the term $(Q, \q)_n$ in vortex partition functions \eqref{genericquivers}. 
 In this section, we revisit this gauging of mass parameter by shifting mass parameters. The key to solve this puzzle is to include the perturbative parts of chiral multiplets.

\subsection{A bi-fundamental puzzle}
To begin with, it is well believed that gauging a fundamental chiral multiplet leads to a bi-fundamental chiral multiplet. In terms of vortex partition functions, a naive observation shows that gauging should be the shift of $x \rightarrow x \q^m$, and then the contribution of each chiral multiplet changes $ (x , \q)_n \rightarrow (x \q^m,\q)_n $ which looks like the contribution from a bi-fundamental matter\footnote{The vortex partition function on Higgs branch can be found in e.g. \cite{Benini:2014aa}.}.
However, this involves a puzzle.
Firstly, the bifundamental matter's contribution take the quiver's form \eqref{genericquivers}, and hence it should be the term
 $(x, \q)_{m+n}$. 
 This term contains three sectors:
\begin{align}
&   (x, \q)_{m +n} = \frac{(x, \q)_\infty}{ ( x  \q^{m+n} , \q )_\infty }  =(x ,\q)_m (x, \q)_n  \frac{  (x \q^m, \q )_n}{(x, \q)_n  }
=(x ,\q)_n (x, \q)_m \frac{  (x \q^n, \q )_m}{(x,\q)_m  } \,.
\end{align}
So the expected bifundamental term $(x \q^m,\q)_n $ does not match with the quiver's form, and there is one more  term that is a fundamental matter. We expect the gauging process preserves the number of matters, while this seems not. This is the puzzle.

If we think of the bifundamental coming from strings between two parallel D3-branes, then these three sectors in $(x,\q)_{m+n}$ are excitations on each branes and the strings stretching between two branes. Therefore, the term  $(x, \q)_{m+n}$ contains two more sectors than the usual bifundamental matter that only contributes $ {  (x \q^m, \q )_n}/{(x, \q)_n  }$ where the denominator removes the case that $m=0$.


Moreover, because of the following identity 
\begin{align}\label{bimn}
(x, \q)_{m+n} =(x, \q)_m (x \q^m, \q)_n  \,,
\end{align}
this term $(x \q^m,\q)_n$ should be viewed as one bi-fundamental matter $(x,\q)_{m +n}$ and one fundamental matter $1/(x,\q)_m$.
Using this identity,
one can further find the origin of $\q$-binomial term is the open strings between two almost overlapped D3-branes:
 \begin{align} \frac{  (x \q^m, \q )_n}{(x, \q)_n  }  =\frac{ (x, \q)_{m +n} }{(x , \q)_m (x, \q)_n }  ~~\xrightarrow{x\rightarrow \q}~~ 
 \frac{ (\q\, \q^{m}, \q)_n }{ (\q,\q)_n }
=
\bigg[
\begin{array}{c}
m+n\\
n
\end{array}
\bigg]_\q   \,.
\end{align} 
Since the bifundamental matter comes from gauging the fundamental matter, we can also think of the $\q$-binomial and open strings arise from gauging mass parameters.

In addition,
this bi-fundamental puzzle may not obvious in the contour integral representation of sphere partition functions, while it emerges in vortex partition functions if we gauge by setting $x =x\q^m$.

\subsection{Gauge mass parameters}

The above puzzle only cause a problem to distinguish either term $(x \q^m,\q)_n$ or $(x, \q)_{m+n}$ to represent the bi-fundamental matter. However, if one takes into account the one-loop contribution, then it becomes clear that the second one should be correct for gauging:
\begin{align}\label{gaugefund}
\frac{ (x , \q)_\inf}{ (x , \q)_n } = (x \, \q^n, \q)_\inf ~~\xrightarrow{x \rightarrow x \q^m}~~\frac{ (x \q^m, \q)_\inf}{ (x \q^m, \q)_n }= (x \q^{m+n}, \q)_\inf = \frac{ (x, \q)_\inf}{ (x, \q)_{m+n} } \,.
\end{align}
Now, we can decide that after gauging the mass parameter, a fundamental matter becomes a bi-fundamental matter:
$$ (x \q^{n}, \q)_\inf =\frac{ (x , \q)_\inf}{ (x , \q)_n }  ~~\xrightarrow{x \rightarrow x \q^m}~~ \frac{ (x, \q)_\inf}{ (x, \q)_{m+n} }  = (x \q^{m+n}, \q)_\inf  \,.$$
If using plumbing graph, the gauging is represented as
\begin{align}\label{guagingmatter}
\bullet_{k_1} -\sqbox  ~~\xlongrightarrow{ \text{gauge mass} }~~ 
\bullet_{k_1}-\sqbox - \bullet_{k_2}
\end{align}
which is consistent with we have observed through sphere partition functions in \cite{Cheng:2023ocj}.
Basically, one can start from gauging a free chiral multiplet:
\begin{align}\label{gaugefreechiral}
 (x , \q)_\inf  ~~\xrightarrow{x \rightarrow x \q^n}~~{ (x \q^n, \q)_\inf}
 = \frac{ (x, \q)_\inf}{ (x, \q)_{n} } \,,
\end{align}
which turns it into a fundamental matter. Further gauging of it leads to a bifundamental matter.

Because of the above analysis, the more appropriate term that denotes a chiral multiplet $\sqbox$ should be $(x,\q)_\inf$.
A naive observation is that no matter how we gauge the mass parameter, the one-loop term is always there and is not affected, as gauging mass parameters only introduces one additional term with a flux number $n$.


In the surgery engineering of 3d theories through three-manifolds $M_3$, there are gauge circles and matter circles. Recall that the former gives gauge groups and the later gives matter fields. Since the gauging that we described above  introduces new gauge groups, we can view gauging as the geometric operation of linking additional gauge circles to matter circles.
For instance, the gauging of a chiral multiplet can be interpreted as 
\begin{equation}\label{fig:massgauge}
\begin{aligned}
\includegraphics[width=2.5in]{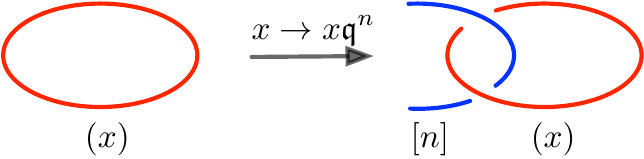}
    \end{aligned}
\end{equation}
where the linking number is the charge $q$ depending on the shift $x \rightarrow x \q^{q n}$.
We assume that each matter circle is attached with a free singlet as its perturbative part, and only when $n=0$, the matter circle only describes the free singlet. Besides, any free singlet can also be given an infinity framing number by doing identical surgery along its neighborhood. When a matter circle $\redcirc$ is linked to a gauge circle $\bluecirc_\inf$ with infinite framing number, then this gauge circle can be deleted freely; see section \ref{sec:STmove} for more details on this.

One can gauge the mass parameter $x$ as many times as possible by recursively using of the identity 
\begin{align} \label{gaugemass12}
(x,\q)_{m+n} = (x,\q)_m (x \q^m,\q)_n \,,
\end{align}
which leads to a multiple charged matter satisfying the formula:
$$ (x,\q)_{d_1 + \cdots + d_k} = (x,\q)_{d_1} (x \q^{d_1},\q)_{d_2} (x \q^{d_1+d_2},\q)_{d_3} \cdots (x \q^{d_1+\cdots +d_{k-1}},\q)_{d_k} \,.$$
Note that a multiple charged matter is linked to many gauge circles and hence carries many flux freedoms from $d_1$ to $d_k$.

From the mass gauging, one can also see that a hypermultiplet $\dbox$ also leads to the $\q$-binormal term:
\begin{align}
\frac{(x, \q)_k}{ (y, \q)_k} ~~\xlongrightarrow{x=x \q^{n}}~~
\frac{ (x ,\q)_{n+k} }{ (y,\q)_k (x,\q)_{n}} ~~\xlongrightarrow{x=y = \q}~~
\begin{bmatrix}
 n+k\\ k
\end{bmatrix}_\q  \,.
\end{align}
One can naively gauge the mass parameter $y=\q \cdot \q^{m}$ to get
\begin{align}
\frac{(y,\q)_k}{(\q,\q)_k} ~~\xrightarrow{ y= \q \cdot \q^{m}}  ~~ \begin{bmatrix}
    m+k \\k
\end{bmatrix}_\q  \,.
\end{align}
This implies that gauging the mass parameter turns a hypermultiplet $\dbox$ into a bi-fundamental chiral multiple $\sqbox$ that arises from the open strings between two gauge nodes $\bullet_k$ and $\bullet_{m}$. 

\subsection{Gauge topological symmetries}
The gauging of topological symmetries is easy, as it involves only the FI parameters in the term $(-\sqrt{\q})^{K_{ij}n_in_i }$ of \eqref{genericquivers}. The geometric interpretation of gauging FI parameters is the Kirby move of the first type, which is also called blow-up/down. In terms of the plumbing graphs, this gauging is denoted by $\bullet_{k}  = \bullet_{k\pm 1}-\bullet_{\pm1}$. After this gauging, a new gauge group is generated; see \cite{Gadde:2013aa,Cheng:2023ocj} for more examples and explanation.

 Gauging the topological symmetry is simply also the replacement of FI parameters by $x \rightarrow x \q^m$, and here $x$ is the FI parameter. This gauging involves the change 
 \begin{align}
\sum_{n=0}^{\inf} (-\sqrt{\q})^{k n^2} x^n  ~~\xrightarrow{x \rightarrow x \q^m}~~ \sum_{n=0}^{\inf} (-\sqrt{\q})^{k n^2+ 2 mn} x^n  \,.
 \end{align}
 In addition, the framing number and the FI parameter of the newly introduce gauge circle can be turned on to get 
  \begin{align}
\sum_{n=0}^{\inf} (-\sqrt{\q})^{k n^2} x^n  ~~\xrightarrow{x \rightarrow x \q^m  ~\oplus ~f ~\oplus~ x'}~~ \sum_{n=0}^{\inf} (-\sqrt{\q})^{k n^2+ 2 mn +f m^2 } x^n x'^m  \,.
 \end{align}
 These can also be turned on for \eqref{fig:massgauge}.
 We can use surgery circles to represent this gauging:
 \begin{equation}\label{fig:FIgauge}
\begin{aligned}
\includegraphics[width=2.5in]{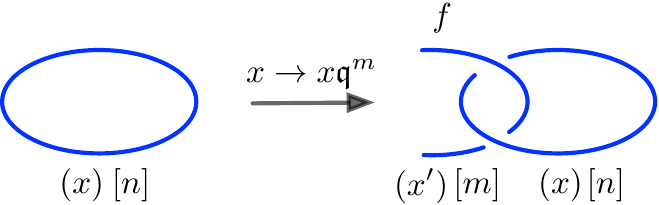}
    \end{aligned}
\end{equation}
Note that the combination of the large mass decoupling limit and the flips in \eqref{variousflips} by sending $x^{\pm1} \rightarrow 0$ is equivalent to a particular gauging of topological symmetries, which can be interpreted as Kirby moves.

\subsection{Quivers arise from gauging}
Let us recall the gauging of a free matter in \eqref{gaugefreechiral},
which turns a free matter into a fundamental matter. The bifundamental itself in \eqref{gaugefund} can be viewed as the gauging through  $x \rightarrow x \q^{m+n}$.
Moreover, for matters with other charges, one can just replace $ m \rightarrow q_1 m\,, n \rightarrow q_2 n $.
The gauging process $x \rightarrow x \q^m$ is to link an introduced gauge circle to a matter circle. In this process, the length of the matter circle as $x$ does not change.

Because of these properties, one can get a new understanding for quiver theories \eqref{genericquivers}, which come from the gauging of a bunch of free matters with background guage fields coupled through mixed CS levels:
\begin{align}
\prod_{\rho=1}^{N_f} (Q_\rho, \q)_\inf ~~\xlongrightarrow{ ~\text{gauging the background}~(K_{ij},\, \xi_i)~ }~~ Z_{K_{ij}}(Q_\rho, \xi_i)  \,.
\end{align}
This interpretation is consistent with the construction of 3d quiver theories by using many unlinked Ooguri-Vafa defects to detect plumbing manifolds.
The gauging is then the process of coupling matter circles to gauge circles, and the latter comes from surgeries.

\section{Flip signs of masses and charges}\label{sec:flips}

Basically, there are three objects that we can flip signs, including mass parameters, charges of matters, orientations of matter circles and orientations of three-manifold. The corresponding flips are $ x \rightarrow x^{-1}$, $n \rightarrow -n$ and $\q \rightarrow \q^{-1}$ respectively. However, the complication is that these flips are not independent.

If only changing the orientation of three-manifolds $\q \rightarrow \q^{-1}$, then equivalently the sign of mass parameter is flipped $x\rightarrow x^{-1}$ as well:
\begin{align}
 (x,\q^{-1})_n = (x^{-1}, \q)_n  (-\sqrt{\q})^{- n^2} (\sqrt{\q} x)^n  \,.
\end{align}
If changing both the orientation $\q \rightarrow \q^{-1}$ and flipping the sign of mass parameter by $x\rightarrow x^{-1}$, then one only get a shift in CS levels and FI parameters:
\begin{align}
(x, \q)_n =( x^{-1}, \q^{-1})_n (-\sqrt{\q})^{n^2} (x/\sqrt{\q})^n \,,
\end{align}
which makes some sense as the mass parameter $x$ is proportional to the volume of the M2-brane wrapping on a holomorphic curve $\mathcal{C}$, namely  $x \sim \exp(\text{vol}(\mathcal{C}) )$, so if the orientation of $\mathcal{C}$ is flipped, then the volume picks up a minus sign.

One can considers other combinations, and the interplay between flipping different parameters are reflected in formulas below:
\begin{align}\label{variousflips}
 &(x, \q)_n =( x^{-1}, \q^{-1})_n (-\sqrt{\q})^{n^2} (x/\sqrt{\q})^n = \frac{(-\sqrt{\q})^{n^2} (x/\sqrt{\q})^n }{ ( \q x^{-1}, \q)_{-n}  }   \,, \\
& (x, \q)_{ -n} = \frac{1}{ (x \q^{-n}, \q)_n} =\frac{1}{ (x \q^{-1}, \q^{-1})_n } = \frac{  (-\sqrt{\q})^{ n^2} ( 
\sqrt{\q} /x)^n }{ (\q x^{-1}, \q)_n }     \,, 
\end{align}
which shows that if flipping only the charge $n \rightarrow -n$, then equivalently either $\q$ or $x$ are flipped. Note that if the orientation of three-manifold is flipped by $\q \rightarrow \q^{-1}$, then everything goes to its anti-version, for instance, branes go to anti-branes. For this case, one can just equivalently flip  mass parameter $x \rightarrow x^{-1}$ and shifts framing numbers.

Note that flipping the orientation of matter circles are also allowed. Since the matter circles come from non-compact three-manifolds, namely the Ooguri-Vafa Lagrangian manifolds in the cotangent bundles of three-manifolds, one can flip only the orientation of the OV defects and preserve that of the base three-manifolds.
An interesting phenomenon is that when the orientations of matter circles are flipped, their contributions to the vortex partition functions go from the denominator to the numerator.

\subsection{Flips in brane webs}

To precisely know how these three flips interacts with each other, we seek help from 3d brane webs. In \cite{Cheng:2021vtq}, the author shows explicit examples and one can conclude that the D5-branes in the following brane webs are responsible to different terms in the vortex partition functions, which are sensitive to the locations of D5-branes.
\begin{equation}
\label{matterbraneweb}
\begin{aligned}
\begin{tikzpicture}
\draw[orange,<->](-1.6, 0.6)--(-1.6,1.4) node[midway, left=-0.1]{\tiny {$\q \leftrightarrow \q^{-1}$}}  ;
\draw[thick,gray] (-1.5,1)--(1.5,1) ;
\draw [line width=1.5pt,red] (-1.5,0.5)--(0,0.5)   (1.5,0.3)--(0,0.3)
  (-1.5,1.5)--(0,1.5) (1.5,1.7)--(0,1.7) ;
  \draw[line width =1.5pt] (0,0)--(0,2) ;
  \draw[line width=3pt, blue] (-0.8,1)--(-1.6,0.2) node[below=0.1,right]{\small{\text{D3}}} ;
  \node at (0,2.2) {\small \text{NS5}}    ;
 \node at ( -0.8,1.75) {\small{\text{D5}}}   ;
 \node at (-2.2, 1.5) {$(a)$}  ;
 \node at (-2.2, 0.5) {$(b)$} ;
  \node at (2.2, 0.3) {$(d)$} ;
    \node at (2.2, 1.7) {$(c)$} ;
    \draw [line width=1.5pt] (-1.6,-0.5)--(-1.6,0.5);
    \node at (-2.1,-0.3)  { \small \text{NS5}'};
    \draw[orange,<->](-0.8, -0.1)--(0.8,-0.1) node[midway, below=-0.1]{\tiny {$\q \leftrightarrow \q^{-1}$}}  ;
\end{tikzpicture}
\end{aligned}
\end{equation}
Where the D3-brane is perpendicular to the plane of the D5-NS5. The strings between D3-D5 is responsible for chiral multiplets. There are four types of open strings stretching from the D3-brane to D5-branes of locations $(a), (b), (c), (d)$, in which  $(a)$ and $(b)$ lead to fundamental chiral multiplets denoted by $\F$ and $(c)$ and $(d)$ lead to  anti-fundamental chiral multiplets denoted by $\AF$ in this note. The contributions of these matters to the vortex partition functions are 
\begin{align}
&&\F: \quad  &(a) ~~\rightarrow ~~ \frac{1}{ (x ,\q)_n}  \,,\quad  &&(b)~~\rightarrow~~ \frac{ 1}{ (\tx, \q^{-1})_{n} } \,, \\
&&\AF: \quad &(c) ~~\rightarrow ~~ (x,\q)_n  = \frac{1}{ ( \q^{-1}x ,\q^{-1})_{-n}} \,,\quad  &&(d)~~\rightarrow~~ (\tx,\q^{-1})_n = \frac{ 1}{ (\q\tx, \q)_{-n} } \,.
\end{align}
One can observe that from $(a)$ to $(c)$ and from $(b)$ to $(d)$, there is a reflection property of $\q \leftrightarrow \q^{-1}$, which from the  perspective of knot invariants is the flip of the orientation. In the above, $x$ and $\tx$ are coordinates involving mass parameters by  $x=e^m$, and $\tx=e^{-\tilde{m}}$ as the $\tx$ is below the original line (the gray line in the figure) of the plane.
In \cite{Cheng:2021vtq}, we already showed that one can reflect the charge of 5-branes from $(p,q)$-branes to $(p, -q)$-branes by the reflection along the original line, which is the flip of the orientations of matters by $\q \rightarrow 1/\q$.

One can notice a cancellation between a fundamental matter and an anti-fundamental matter $1\F +1 \AF \rightarrow \emptyset $, which is also a particular Higgsing or conifold transition. In terms of partition functions, this cancellation is trivial
$ \frac{1}{(x, \q)_n} \cdot {(x, \q)_n} =1$\,,
however this is not too trivial if we rewrite the contribution of the antifundamental matter through the identity
\begin{align}\label{rewritefund}
(x, \q)_n =\frac{1}{(\q^{-1}x ,\q^{-1})_{-n}} \,,  
\end{align}
which means that if a $\F$ has an opposite orientation and an opposite charge in comparison with another $\F$, they can be cancelled because of
\begin{align}\label{cancelbraneanti}
\frac{1}{ (x,\q)_n  (\q^{-1}x ,\q^{-1})_{-n}  }  =1 
\,,
\end{align}
which means that in other words, if we flips both the charge and the orientation of the M2-brane, then we get an anti-D5-brane which cancels the D5-brane.
In addition,
for an unknot in three-manifolds, we have an associated  hypermultiplet that is equivalent to one $\F$ and one $\AF$, and hence we have the following associated terms in vortex partition functions 
\begin{align}\label{unknotmatter}
 \frac{1}{ (x, \q)_n {(\q^{-1} y ,\q^{-1})_{-n}}  } \,,
 \end{align}

Because of the following identity, one can see that a pair of brane and anti-brane (the flipped D5-brane) only differ by a Chern-Simons level and a FI parameter:
\begin{align}\label{equivabrananti}
\boxed{
{(x, \q)_n} = {(x^{-1}, \q^{-1})_n } \cdot {(-\sqrt{\q})^{n^2} (x/\sqrt{\q})^n} }   \,.
\end{align}
One can then consider if there is a cancellation between a fundamental matter $(a)$ and a flipped anti-fundamental matter $(d)$, the answer is that they are almost cancelled:
\begin{align}\label{braneandantibrane}
\boxed{
\frac{1}{(x ,\q)_n (\q x^{-1}, \q)_{-n}} = \frac{1}{(-\sqrt{\q})^{n^2} (x/\sqrt{\q})^n }  }
\,.
\end{align}
where we have used \eqref{rewritefund}, and one can more easily see this cancellation from \eqref{equivabrananti}. This almost cancellation can also happen between $(c)$ and $(d)$ if we flip $\q$.
This almost cancellation means a brane almost cancels with a brane with an opposite charge and mass parameter.

\subsection{Flips as twists of three-manifolds}

Having clarified the relations between different matters, we can compare with three-manifolds and Ooguri-Vafa defects to find a geometric realization of these properties.

The almost cancellation \eqref{braneandantibrane} indicates that a pair of brane and anti-brane emerges from a pure gauge node $\bullet_{\pm 1}$ where there are two cases $\pm1$ because of \eqref{rewritefund}. Note that we should loose the restriction that mass parameters for both $\F$ and $\AF$ are $x$ to let them deform independently, such that the brane and the anti-brane could emerge.

In the following, we can interpret the flip from  \eqref{cancelbraneanti} to  \eqref{braneandantibrane} in the following figure: 
\begin{equation}\label{fig:flipmass}
\begin{aligned}
\includegraphics[width=2.8in]{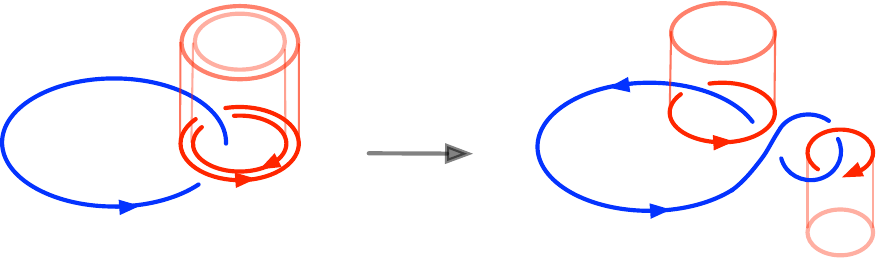}
    \end{aligned}
\end{equation}
where we only draw the M2-branes which are the red tubes, and do not draw the Ooguri-Vafa (OV) defects that one boundary of M2-brane ends on for simplicity. Note that these OV defects have the topology of the solid torus $\mathbb{R}^2 \times \redcirc$. The left graph, namely \eqref{cancelbraneanti} describes the emergence of a pair of M2-brane  and its anti-brane, and one can deform them by giving them different mass parameters (lengths of red circles). Note that the orientations of this pair of branes are opposite. We can rotate one red tube from the upward to the downward, then the sign of its mass parameter is flipped and the orientation is also flipped, but the charge as the winding number preserves; see \eqref{variousflips}. Due to this rotation, the gauge circle (blue circle) is twisted and picks up an additional $\pm 1$ framing number. In addition, one can take the massless limit by  shrinking the length of the red tube as the mass parameter $x \sim \exp(-\text{length of } \redcirc )$. The decoupling limit is given by taking the height of the tube to be infinite large. The decoupling does not change the framing number of the linked gauge circles. 

Moreover, matters with generic charged matters by replacing $n \rightarrow \sum q_i n_i$ can also be described by a twist in the graph in \eqref{fig:flipmass}. If we flip a bifundamental matter, then the introduced twist is $(n_1+n_2)^2= n_1^2 +n_2^2+2n_1n_2$ which causes a linking between gauge circles $\bullet_{[n_1]}-\bullet_{[n_2]}$ where the subscripts denote their flux numbers. In this generic case, the flips are just the Fenn-Rourke moves which are equivalent to a Kirby moves of the first type, but still we need to keep the matter circles, so these Fenn-Rourke moves here are decorated because of the presence of matter circles. The original Fenn-Rourke move is illustrated in below:
\begin{equation}\label{fig:Fenn}
\begin{aligned}
\includegraphics[width=1.5in]{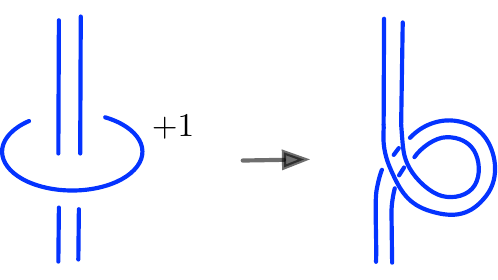}
    \end{aligned}
\end{equation}
See e.g.\cite{Rolfsenbook,Prasolovbook} for more details on this move.
This move is very useful as it is believed that for three-manifolds the combinations of Fenn-Rourke move and handle slides that we will discuss later give a complete set of Kirby moves.

\section{ST-duality}\label{sec:STmove}
In this section, we discuss a basic 3d duality as an example to show the crucial roles played by gauging and flips.

As we have mentioned before, ST-move is the gauged version of the ST-duality, which is named for the duality between a free chiral multiplet $\Phi$ and a theory $U(1)_{\pm 1/2} + 1 \Phi$ where the matter is in the fundamental representation. The ST-dualtiy is found in \cite{Dimofte:2011ju}, and is denoted as $\sqbox ~\leftrightarrow~ \bullet_{\pm 1/2} -\sqbox$. If we use the effective Chern-Simons levels, the ST-duality is denoted by $\sqbox ~\leftrightarrow~ \bullet_{\pm 1} -\sqbox$. ST-move is obtained by gauging the topological symmetry and the mirror dual flavor symmetry to get
$ \bullet_k -_q\sqbox ~\leftrightarrow~ \bullet_{k\pm q^2}-_q\bullet_{\pm1}-\sqbox$ \footnote{where $q$ in $-_q$ is the charge of $\sqbox$, or the linking number if it is between two gauge nodes, such as $\bullet -_q\bullet$.}. ST-move is discussed in \cite{Cheng:2023ocj}. 
 
In this note, we add the detail that 
if a gauge node $\bullet_\inf$ links to a matter node $\sqbox$, then this gauge node $\bullet_\inf$ can be ignored. We can use ST-move to draw this conclusion:
\begin{equation}\label{STmovegraph}
\begin{aligned}
\includegraphics[width=2.8in]{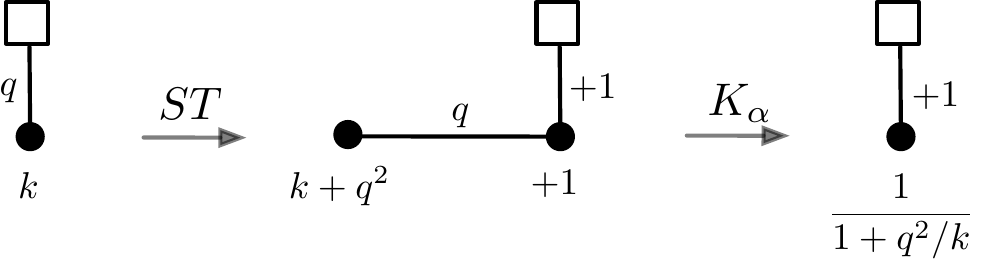} 
    \end{aligned}
\end{equation}
where in the second step we apply the Kirby move to integrate out $\bullet_{k+q^2}$ to get the third graph.
When $k=\inf$, the third graph becomes $\bullet_{+1}-\sqbox = \sqbox$ which equals to the first graph $\bullet_\inf-{\sqbox} $.
This means that when a node $\bullet_\inf$ attaches to a $\sqbox$, this node $\bullet_\inf$ can be freely deleted.
In particular, we remind a special case that when $q=1$ the first graph relates to the third graph in \eqref{triangle} by a Rolfsen's twist, which is an equivalent surgery. This confirms the geometric realization that ST-duality is Rolfsen's twist \cite{Cheng:2023zai,Rolfesn:1984}.

We want to find out an identity for quiver/vortex partition functions to represent the ST-duality. Although there are many well known identities for $\q$-Pochhammer products, we do not clearly know which one is for ST-duality. The author already did computations for sphere partition functions, but these identities for sphere partition functions are contour integrals and are not obvious to show BPS invariants. A naive solution is that we can observe some basic $\q$-series identities and looks at the associated effective Chern-Simons levels $K_{ij}$ to match with the plumbing graphs for ST-dualities and ST-moves.

\vspace{2mm}
\noindent
\textbf{$(ST)^2$-move.}
The most basic identity is
\begin{align}\label{STreverse}
\boxed{
\frac{1}{ (x,\q)_\inf } =\sum_{n=0}^\inf  \frac{ x^n}{(\q,\q)_n} }  \,.
\end{align}
This identity however has a vanishing CS level, and the associated plumbing graph is $\sqbox ~\leftrightarrow~ \bullet_0 -\sqbox$. As is shown in \cite{Cheng:2023ocj}, we can known that this dual graph is obtained by performing $ST$-moves two times and then integrating out the gauge node in the middle. 
This duality can also come from the decoupling limit $y \rightarrow 0$ of the S-dual pair \eqref{inffusion}.
We can gauge this dual pair by $x =x \q^{{ q} m}$ to get
\begin{align}\label{ST2move}
{
\frac{(x,\q)_{q m}}{ (x, \q)_\inf } =\sum_{n=0}^\inf  \frac{x^n}{(\q,\q)_n} \cdot   (-\sqrt{\q})^{2 q \, m n}  \,.
}
\end{align}
This duality turns a matter with charge $-q$ into a matter with charge $+1$. The associated plumbing graph is $ \bullet_k-_{-q}\sqbox ~\leftrightarrow~\bullet_k-_q\bullet_0-\sqbox $ where we have turned on the framing number $k$. 
We remind that the presence of the one loop term $(x,\q)_\inf$ is necessary for further gauging $\tx$ to get bifundamentals; see \eqref{gaugefund}. If one further turns on the degree of freedom for $m$ on both sides, namely the gauge node $\bullet_k$, the FI parameter and CS levels can be turned on by adding $\sum_{m=0}^{\inf} (-\sqrt{\q})^{k\,  m^2 } \xi^{m}$ on both sides. 

If flipping the mass parameter and charge using the identity \eqref{variousflips} in section \ref{sec:flips} to use the quiver form in \eqref{genericquivers}, one gets
\begin{align}
\frac{(\q x^{-1}, \q)_\inf }{ \theta(-x /\sqrt{\q} ) } \cdot \frac{ (-\sqrt{\q})^{q^2 m^2} (x/\sqrt{\q})^{ q m}
} {(\q x^{-1}, \q)_{-qm}}=\sum_{n=0}^\inf  \frac{x^n  }{(\q,\q)_n} \cdot  (-\sqrt{\q})^{ 2 q\, mn}  \,,
\end{align}
where $ \theta(-x /\sqrt{\q} )  = (x, \q)_\inf  (\q x^{-1}, \q)_\inf $. The associated plumbing graph for this move is $ \bullet_{k+q^2} -_{\tiny{-q}}\sqbox ~\leftrightarrow~ \bullet_{k}-_{q} \bullet_0-\sqbox$, which obviously cannot be a Kirby move when we decouple $\sqbox$, but still has geometric meaning in terms of three-manifolds if we taking into account the twist in \eqref{fig:flipmass}. Notice that a theta function appears which in principle involves the contribution from the theory on the boundary of the 3d theory \cite{Gaiotto:2008ak,Beem:2012mb}, which deserves a further discussion in future.

\vspace{2mm}
\noindent
\textbf{$ST$-move.}
The correct identity for $\bullet_{+1}-\sqbox$ actually arises from the twisted S-duality \eqref{fundtobi2} in the limit $n \rightarrow \inf$. This flipped decoupling $y^{-1} \rightarrow 0$ is an appropriate decoupling of the S-dual pair \eqref{inffusion}, from which one can see a descendent identity:
\begin{align}\label{STduality}
\boxed{
(\tx, \q)_\inf  = \sum_{n=0}^\inf 
\frac{ ( \tx/\sqrt{\q} )^n  }{ (\q,\q)_n }  \cdot ( - \sqrt{\q})^{n^2}   } \,,
 \end{align}  
where $\tx = x y$ and we have used \eqref{variousflips}. The plumbing graph for this identity is $\sqbox ~\leftrightarrow~ \bullet_{+1} -\sqbox$, hence \eqref{STduality} looks perfectly fine for the ST-duality.

If gauging the twisted $\tx =\tx \q^{q m}$ without flips, one can get a gauged duality:
\begin{align}\label{STmovequiver}
\boxed{
\frac{ (\tx, \q)_\inf}{(\tx, \q)_{q m }} = \sum_{n=0}^\inf 
\frac{ ( \tx/\sqrt{\q} )^n  }{ (\q,\q)_n }  \cdot ( - \sqrt{\q})^{n^2+2 q \, m n} }  \,,
\end{align} 
which describes the move $\bullet_{k}-_q \sqbox ~\leftrightarrow~ \bullet_{k}-_q\bullet_{+1}-\sqbox$.
The above identity for partition functions agrees with the decoupling of D5-branes in 3d brane webs that are discussed in \cite{Cheng:2021vtq}.
Briefly, this duality as decoupling a half of D5-brane of the brane web in \eqref{u1faf} can be partially illustrated as follows:
\begin{equation}
\begin{aligned}
\includegraphics[width=3in]{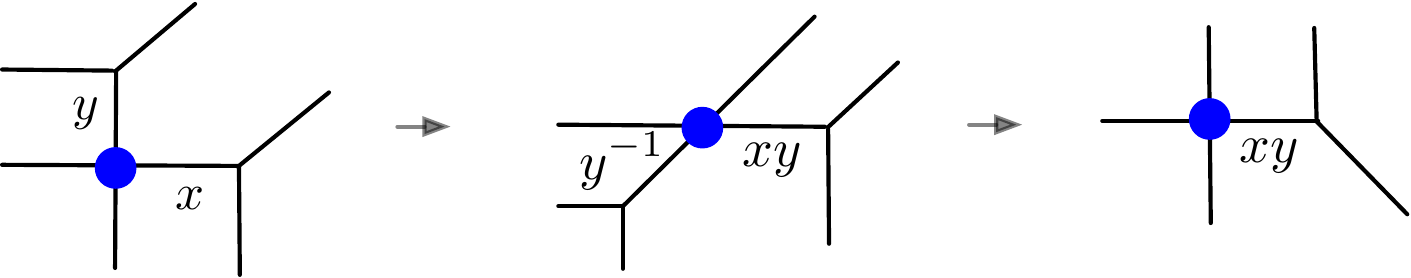}
    \end{aligned}
\end{equation}
from which one can see how the parameter $\tx =xy$ emerges.

If one wants to match ST-moves with Kirby moves at an appropriate decoupling limit of matters and hence a good geometric realization, the signs of the mass parameters on both sides should be opposite.
Using once again the formula \eqref{variousflips}
we can flip the sign of the mass parameter on the LHS of \eqref{STmovequiver}, and hence get
\begin{align}\label{STflipmass}
 \theta(-\tx/\sqrt{q}) \cdot  \frac{ ( \q \tx^{-1}, \q)_{-qm}  } 
{(-\sqrt{\q})^{q^2 m^2} (\tx/\sqrt{\q})^{q m} }= \sum_{n=0}^\inf 
\frac{ ( \tx/\sqrt{\q} )^n  }{ (\q,\q)_n }  \cdot ( - \sqrt{\q})^{n^2+2 q \, m n}   \,.
\end{align} 
The associated plumbing graph is $\bullet_{k-q^2}-_q \sqbox ~\leftrightarrow~ \bullet_{k}-_{q} \bullet_{+1}-\sqbox$ which looks able to match with Kirby move of the first type that is $\bullet_{k-q^2} ~\leftrightarrow~ \bullet_k-_q \bullet_{+1}$. 

This example tells that we should perform flips or twists in \eqref{fig:flipmass} to nicely realize the ST-move as an extended Kirby move.
Here, the flip of the sign causes significant changes to the identity, which even moves the $\q$-Pochhammer product from the denominator to numerator. This corresponds to a geometric change of the position of the M2-brane as a tube from upward to downward, as is shown in the figure in \eqref{fig:flipmass}. This geometric change can be observed through vortex partition functions. The introduced twist leads to the shift of framing number by $q^2$; see textbook \cite{Rolfsenbook} for this detail on twists.

In terms of surgeries and Kirby moves, one can understand the consistency of ST-moves.
\begin{equation}\label{RtwistKa}
\begin{aligned}
\includegraphics[width=3.5in]{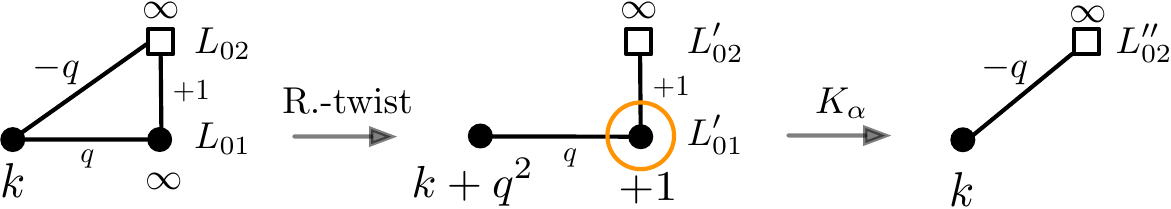}
    \end{aligned}
\end{equation}
where we already know that the first graph is equivalent to the third graph as the $\bullet_\inf$ connecting to $\sqbox$ can also be deleted freely. From the first graph to the second, we just performed the Rolfsen's twist \cite{Rolfesn:1984} which is an equivalent surgery that is particularly applied to $\bullet_\inf$ as its framing number is $\inf$ the usually way is not easy to analyze it. For the second graph to the third graph, what we did is just the Kirby move of blowing down the node $\bullet_{+1}$ although we have not figured out if we can integrate out a gauge node connecting to a matter node, which however leads to a negative charge which is opposite to the ST-move that tells it should be a positive charge, see \eqref{STmovegraph}. This is a puzzle but can be justified if we consider the flip of mass parameters as the twist illustrated in \eqref{fig:flipmass} and \eqref{variousflips}, which is a hiding movement of M2-branes that has not been reflected in our matter node $\sqbox$. In addition, Kirby moves of the first type are Fenn-Rourke moves. We should give a sign $\sqbox_{<,>}$ to denote if the M2-brane have upward or downward orientations. However, in this note we will not go to these very detailed differences.

\section{Handle slides for gauge circles}\label{sec:handleslides}
Handle slides as the Kirby move of the second type are always considered as equivalent surgeries in constructing invariants for three-manifolds and knots \cite{kirbymove}. 
We noticed there are two kinds of handle slides  as there are two types of circles: gauge circles and matter circles. In literature, the handle slides for gauge circles are often checked whenever constructing new three-manifolds invariants and have become a necessary condition. In this section, we will only show that the handle slides for gauge circles are crucial in many aspects, as these operations transform plumbing graphs significantly in the presence of matter nodes.


To begin with,
the handle sliding has a very simple interpretation, which is the linear recombination of surgery circles
\begin{align}
\{ L_i \} ~~\xlongrightarrow{K_\beta}~~ \{ \tL_i \}  \,,
\end{align}
where each $L_i$ denotes a surgery circle with a framing number as the self-linking number $f_i= L_i \cdot L_i$. Noticing that these $L_i$ is the longitude of the torus boundary of each link complement in the complement $S^3 \backslash (D^2 \times L_i)$. 

In the presence of matter, we denote the handle slide by $K_\beta$, which transforms that plumbing graphs as follows 
\begin{equation}
\label{slidlesmatter}
\begin{aligned}
\includegraphics[width=3in]{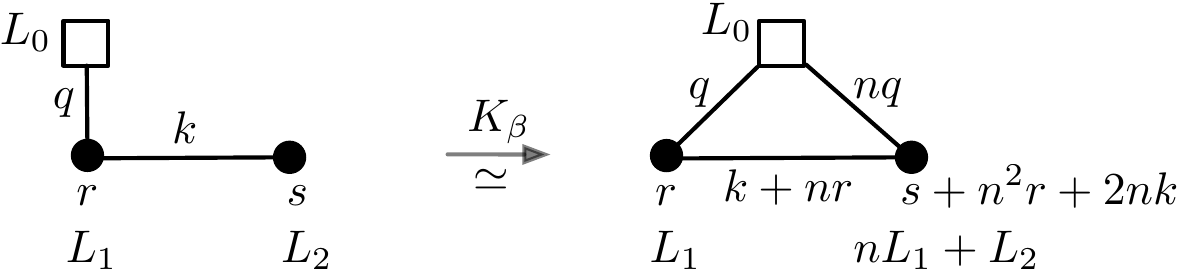}
    \end{aligned}
\end{equation}
where we have assigned the recombination of surgery circles on the graph, and the matter circle $L_0$ is also a surgery circle that links to the recombined circle $nL_1+L_2$ after handle slides. It is easy to use ST-move on $\sqbox$ and Kirby move $K_\a$ to integrate out node $\bullet_r$ to prove \eqref{slidlesmatter}.
Notice that since the matter circle $L_0$ has an infinity framing number, it would be a disaster if combining it with a surgery circle of a finite framing number, so we avoid address $\sqbox$ directly at this stage.

It is better to give details on handle slides. Firstly,
the recombination of surgery circles are 
\begin{align}\label{slidecircles}
    \begin{bmatrix}
        \widetilde{L}_1 \\ \widetilde{L}_2 
    \end{bmatrix} =
     \begin{bmatrix}
    1 & 0 \\ n & 1
    \end{bmatrix}
     \begin{bmatrix}
        {L}_1 \\ {L}_2 
    \end{bmatrix} \,.
\end{align}
The transformation matrix is an element in the mapping class group $SL(2,\mathbb{Z})$.
Since FI parameters are associated with surgery circles, the new FI parameters follow the same transformation
\begin{align}
    \begin{bmatrix}
        \widetilde{\xi}_1 \\ \widetilde{\xi}_2 
    \end{bmatrix} =
     \begin{bmatrix}
    1 & 0 \\ n & 1
    \end{bmatrix}
     \begin{bmatrix}
        {\xi}_1 \\ {\xi}_2 
    \end{bmatrix} \,.
\end{align}

We should consider the effect of handle slides in the path integral.
Through localization, the path integral becomes a contour integral over scalar fields $\oint d \phi_1 d\phi_2 e^{\mathcal{L}(\phi_i)}$. Because the FI parameters  as the background fields are paired with scalars $\xi_i \phi_i$, transforming FI parameters requires to transforming the scalar fields in vector multiplets by the following way:
\begin{align}
    \begin{bmatrix}
{\phi}_1 \\ {\phi}_2 
    \end{bmatrix} =
     \begin{bmatrix}
    1 & n \\ 0 & 1
    \end{bmatrix}
     \begin{bmatrix}
        \widetilde{\phi}_1 \\ \widetilde{\phi}_2
    \end{bmatrix} \,.
\end{align}
The magnetic flux numbers are associated with surgery circles. The flux numbers $d_i$ are degrees of FI parameter through terms $x_i^{d_i} $ where $x_i=e^{\xi_i}$ and correspondingly $d_i$ should changes in the same way as $\phi_i$,
\begin{align}\label{slideflux}
    \begin{bmatrix}
{d}_1 \\ {d}_2 
    \end{bmatrix} =
     \begin{bmatrix}
    1 & n \\ 0 & 1
    \end{bmatrix}
     \begin{bmatrix}
        \widetilde{d}_1 \\ \widetilde{d}_2
    \end{bmatrix} \,.
\end{align}

Because of these tricky changes of parameters, it may take a few minutes to recognize if an identity is caused by a handle slide. It is more better to firstly keep track of the linking numbers of surgery circles and then identify FI and mass parameters; see \eqref{STmoveslidesex1} for an example on how these parameters change under handle slides. 

In terms of 3-manifolds and surgery, the handle slides of gauge circles are geometrically represented as
\begin{equation}
\begin{aligned}
\includegraphics[width=3in]{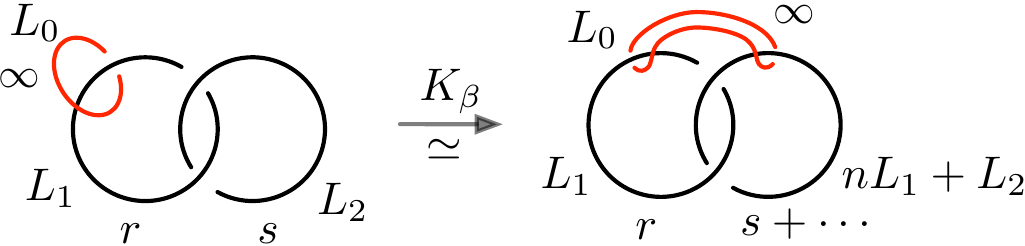}
    \end{aligned}
\end{equation}
If the circles $L_1$ and $L_2$ is linked to external circles, then correspondingly the linking graph will change after handle slides. For instance, if there is an external circle $L_3$ linking to $L_1$, then it will links to the $nL_1 +L_2$ as well. If the $L_1$ and $L_2$ are associated with FI parameters $x_1=\exp(\xi_1)$ and $x_2 =\exp (\xi_2)$, then after this handle slides, the FI parameters become $x_1$ and $x_1^n x_2 =\exp (n\xi_1 + \xi_2)$.

In particular, if $k=0$ and $r=0$, then the handle slides is simple, as there is no linking number generated:
\begin{equation}\label{slidezerofund}
\begin{aligned}
\includegraphics[width=4.5in]{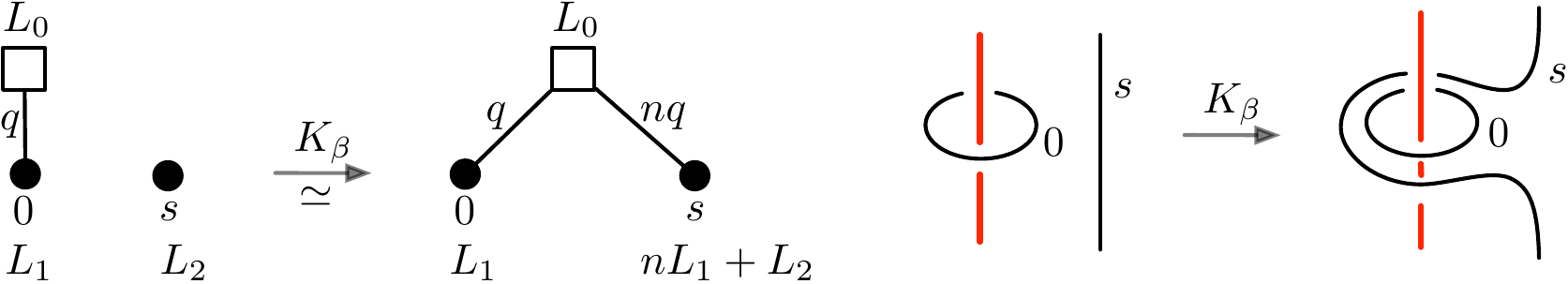}
    \end{aligned}
\end{equation}
In this case, although it looks that the $\bullet_0-_q\sqbox$ connects to $\bullet_s$, but equivalently, this component decouples.

One can combine ST-moves and handle slides to get
\begin{equation}\label{slideST}
\begin{aligned}
\includegraphics[width=4in]{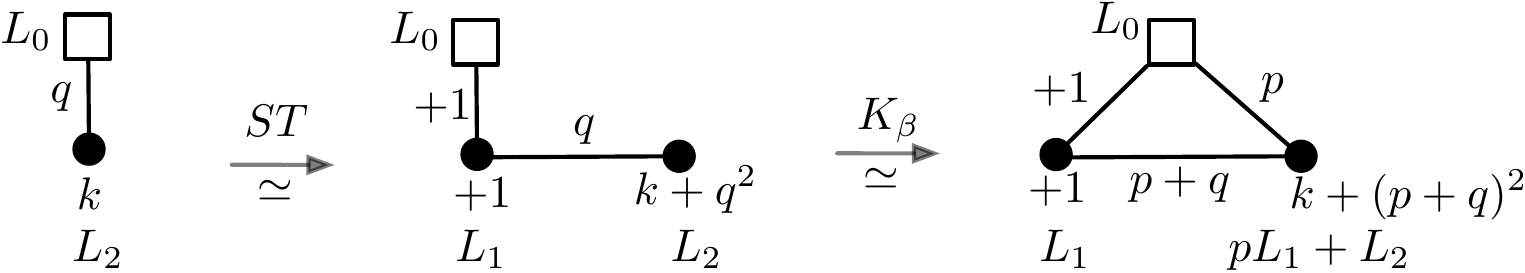}
    \end{aligned}
\end{equation}
In particular, when $p=-q$, the above graph reduces to 
\begin{equation}
\begin{aligned}
\includegraphics[width=2.5in]{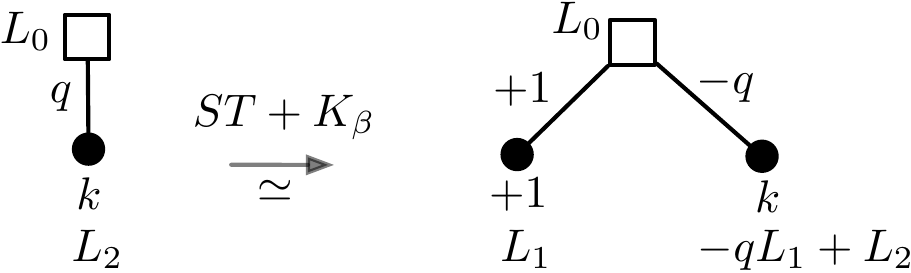}
    \end{aligned}
\end{equation}
which looks a special gauging of mass parameter with the difference that the charge flips sign.

Another special case is that when $q=0$, which leads the coupling of a free matter to gauge nodes:
\begin{equation}
\begin{aligned}
\includegraphics[width=3.5in]{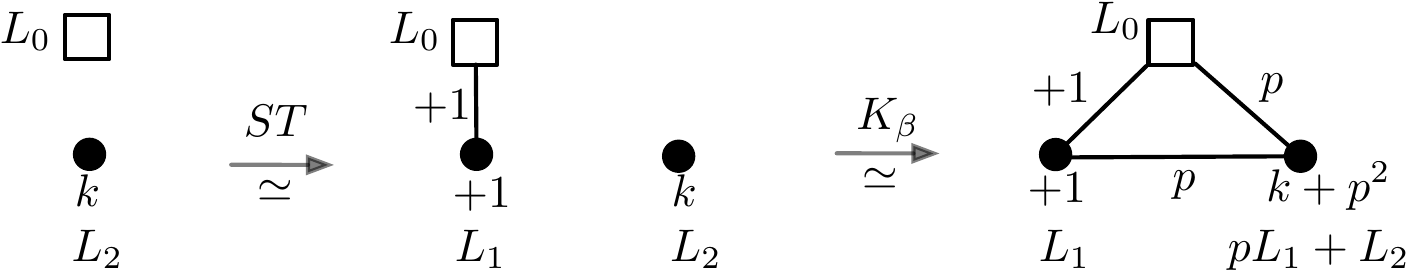}
    \end{aligned}
    \end{equation}
    One can $(ST)^{2}$-move the $\sqbox$ in the third graph to go back to the first graph. Analogously, one can use handle slides to equivalently couple to many gauge nodes, although this coupling seems only increase complications.

\vspace{4mm}\noindent
\textbf{$\q$-binomials.}
The $\q$-binomial terms often appear in the knot invariants and quantum groups. Fortunately, they can be reorganized into simple forms combinatorially. Here, we will show that these combinatorially combinations are just handle slides. 

Let us use an example to explain.
The expansion of two products is
\begin{align}\label{qbinormalex1}
\begin{bmatrix}
    k_1+k_2+k_3 \\ k_2 + k_3
\end{bmatrix}
\begin{bmatrix}
    k_2+k_3 \\ k_3 
\end{bmatrix} = \frac{(\q,\q)_{k_1+k_2+k_3} }{ (\q,\q)_{k_1} (\q,\q)_{k_2} (\q,\q)_{k_3} }
\end{align}
which contains a tri-fundamental matter and three fundamental matters. One can perform handle slides and redefine $\tk_i$ to rewrite the above expression as 
\begin{align}\label{handleex111}
\begin{bmatrix}
    \tilde{k}_1 \\ \tk_2 
\end{bmatrix}
\begin{bmatrix}
    \tk_2\\ \tk_3 
\end{bmatrix} = \frac{(\q,\q)_{\tk_1} }{ (\q,\q)_{\tk_1-\tk_2} (\q,\q)_{\tk_2-\tk_3} (\q,\q)_{\tk_3} }  \,,
\end{align}
which contains two bi-fundamental matters, one fundamental matter, and one anti-fundamental matter. There is no need to worry about the constraints that $k_i > k_{i+1}$ that are caused handle slides, as this constraint is not visible in other contexts, such as the path integrals of sphere partition functions. This may weak the advantage of vortex partition functions, but does not reduce its ability to detect the geometry. We expect using holomorphic blocks in \cite{Beem:2012mb} could fill this gap. 

 One can use plumbing graphs to represent the above $\q$-binomial products in \eqref{qbinormalex1} and \eqref{handleex111}:
\begin{equation}
\begin{aligned}
\includegraphics[width=3.5in]{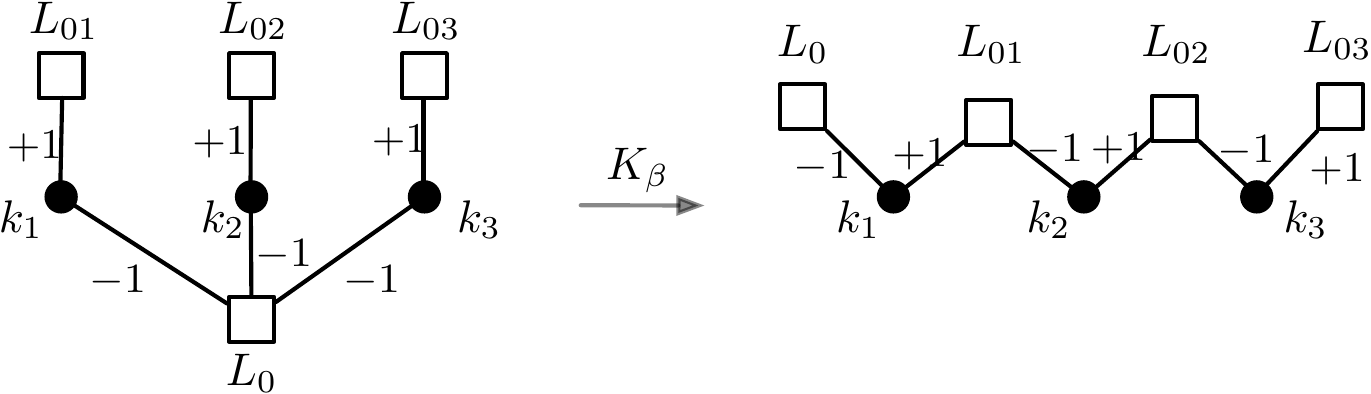}
    \end{aligned}
\end{equation}
where the handle slide is
\begin{equation}\label{handletri}
\{ L_1 \,, L_2\,, L_3  \} ~~\xrightarrow{K_\beta}~~ 
\{L_1\,, L_2- L_1\,, L_3-L_2    \} \,.
\end{equation}
Note that the shifts of circles $L_i$ is not the same as that of degrees $\tk_i$, but reversely, as is shown in \eqref{slidecircles} and \eqref{slideflux}.
Note that the tri-fundamental on the bottom of the left graph is moved to the first node of the right graph.

 In the analysis of HOMFLY-PT polynomials for KQ correspondence \cite{Kucharski:2017ogk}, the term $x^r/(\q,\q)_r$ is added, which could remove the term $(\q,\q)_r$ in the denominator, which is a brane and anti-brane cancellation. Basically,
\begin{equation}
    P(x) = \sum_{r=0}^{\inf} \frac{ x^r}{ (\q,\q)_r} P_r(a,\q) =\sum_{r=0}^{\inf} \frac{ x^r}{ (\q,\q)_r} \sum_{d_1+d_2+\cdots+d_m=r} \frac{(\q,\q)_r \cdot \text{other terms}}{ (\q,\q)_{d_1}(\q,\q)_{d_2} \cdots (\q,\q)_{d_m} } \,.
\end{equation}
One can draw the plumbing graph to represent this partition funciton
\begin{equation}
\begin{aligned}
\includegraphics[width=1.2in]{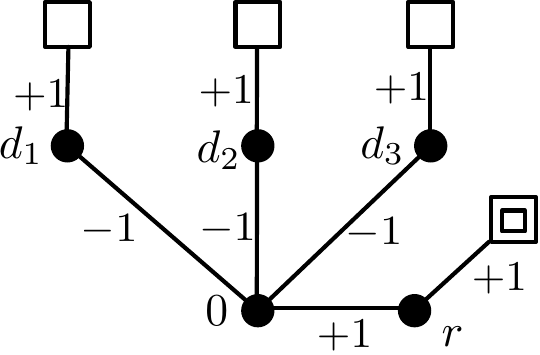}
    \end{aligned}
\end{equation}
where the $\dbox$ is a pair of $\F$ and $\AF$, which cancel each other, and hence $\dbox$ can be ignored and only a  gauge node $\bullet_0$ left. 
The relation $d_1+d_2+\cdots+d_m=r$ can be understand through the above example, and it is just a bi-product of handle slides. 

From the above examples, we notice that the handle slides for gauge nodes could significantly change the forms of the plumbing graphs and quiver partition functions, although it is just the recombination of scalar fields for abelian 3d theories.

\section{Fusions for matter circles}\label{sec:fusion}
In this section, we discuss a particular handle slide for matter circles, which can be viewed as the connected sum of matter circles. This connected sum is described by a fusion identity. We also discuss its children by taking various degeneracy limits.

\subsection{Fusion identity}

There is an identity that had been written repeatedly, see \cite{George} for a derivation. This identity is 
\begin{align}\label{fusion}
\boxed{ 
\frac{ (x y, \q)_n }{ (\q, \q)_n } =
\sum_{k=0}^n \frac{ (x ,\q)_{n-k }}{ (\q,\q)_{n-k} } \cdot \frac{ (y,\q)_k}{(\q,\q)_k} \cdot x^k    }  \,,
\end{align} 
we name it fusion identity and will justify this name later.
It is easy to check that the fusion identity does not pick up any additional term and its form is preserved if flipping the orientation by $\q \leftrightarrow \q^{-1}$ and exchanging $x$ and $y$:
\begin{align}\label{qbinormalex11}
\frac{ (x^{-1} y^{-1}, \q^{-1})_n }{ (\q^{-1}, \q^{-1})_n } =
\sum_{k=0}^n \frac{ (x^{-1} ,\q^{-1})_{n-k }}{ (\q^{-1},\q^{-1})_{n-k} } \cdot  (y^{-1})^{n-k}  \cdot \frac{ (y^{-1},\q^{-1})_k}{(\q^{-1},\q^{-1})_k}   \,.
\end{align} 
One can see that mass parameters and FI parameters flip signs, so the orientations of the associated surgery and gauge circles flip $\{L_i \}\rightarrow \{-L_i \}$. Although the positions of OV defects did not change, but their orientations are flipped, because the orientation of the base three-manifold is flipped: $M_3 \leftrightarrow -M_3$.

Another observation is that in fusion identity the mass parameter in  $(x,\q)_{n-k}$ equals to the FI parameter in $x^k$. This is a puzzle and we have not figured out what this means in 3d theories.

The fusion identity can be rewritten as
\begin{align}
(xy, \q)_n = \sum_{k=0}^n  \bigg[
\begin{matrix}
    n\\k
\end{matrix} \bigg]_\q
 (x,\q)_{n-k} \cdot (y,\q)_k x^k  \,,
\end{align}
which means that three open strings with lengths $x$, $y$ and $\q$ are joined into an open string with length $xy$. Another form is
\begin{align}
\frac{(xy, \q)_n}{ (x,\q)_n} = \sum_{k=0}^n  \bigg[
\begin{matrix}
    n\\k
\end{matrix} \bigg]_\q \frac{(y,\q)_k}{(x \q^{n-k},\q)_{k}} \cdot x^k  \,,
\end{align}
which often appears in knot invariants, but we have not found an intuitive interpretation for this form.

One can try to understand the fusion identity from the perspective of gauging mass parameters of free singlets $(\cdot, \q)_\inf$. The term $(xy, \q)_n$ could arise from the gauging $\G_3: xy \mapsto xy \q^n$, and $(x,\q)_{n-k}$ and $(y,\q)_k$ come from $\G_1: x \mapsto x \q^{n-k}$ and $\G_2: y\mapsto y \q^k$ respectively. Obviously, we have a flux number conservation that $\G_3 =\G_1 \circ \G_2$. Hence the flux number $k$ is a middle process and should be integrated out by summation.

\vspace{2mm}
\noindent
\textbf{Handle sliding gauge circles in fusion identity.} 
The fusion identity involves two gauge symmetries with flux numbers $n$ and $k$. As we have discussed in section \ref{sec:handleslides}, we can recombine them to perform handle slides, which do not change the three-manifolds themselves.

If redefining $d=n-d_1$ with $d_1=k$ and turning on framing number $f$, the fusion identity leads to
\begin{align}\label{slideexf}
\sum_{n=0}^\inf  (-\sqrt{\q})^{ f n^2}\frac{ (x y, \q)_n }{ (\q, \q)_n } 
z^n
=\sum_{d=0}^\inf
 (-\sqrt{\q})^{ f d^2  } \frac{ (x ,\q)_{d }}{ (\q,\q)_{d} } z^d
\cdot (-\sqrt{\q})^{ 2 f d d_1  }  \cdot
\sum_{d_1=0}^\inf
(-\sqrt{\q})^{ f d_1^2 } 
\frac{ (y,\q)_{d_1}}{(\q,\q)_{d_1}} (xz)^{d_1} \,,
\end{align}
for which the surgery circles are changed by the handle slide: $$ \{L_1, L_2\} \rightarrow \{ L_1, L_1+L_2   \} $$ where $L_1$ and $L_2$ are associated with gauge nodes $\bullet_{[n]}$ and $\bullet_{[k]}$ respectively.

When the framing number $f=0$, by using the identity \eqref{inffusion}, one can find the equivalence \eqref{slideexf} is transformed into a trivial identity $ (xyz,\q)_\inf/(z,\q)_\inf = (xz,\q)_\inf/(z,\q)_\inf \cdot (yxz,\q)_\inf/(xz,\q)_\inf$.
It is interesting that in \cite{George} this trivial identity is used to derive the fusion identity by sliding back to $n$ and $k$, and throwing away the sum over $n$. This implies the closed relation between S-dual pair \eqref{inffusion} and the fusion identity \eqref{fusion}, as they can derive each other, but still not equivalent.

\vspace{2mm}
\noindent
\textbf{Fusion chains.} 
One can continue applying the fusion process for infinite many times:
\begin{equation}\label{fusionchain}
\begin{aligned}
\includegraphics[width=4.5in]{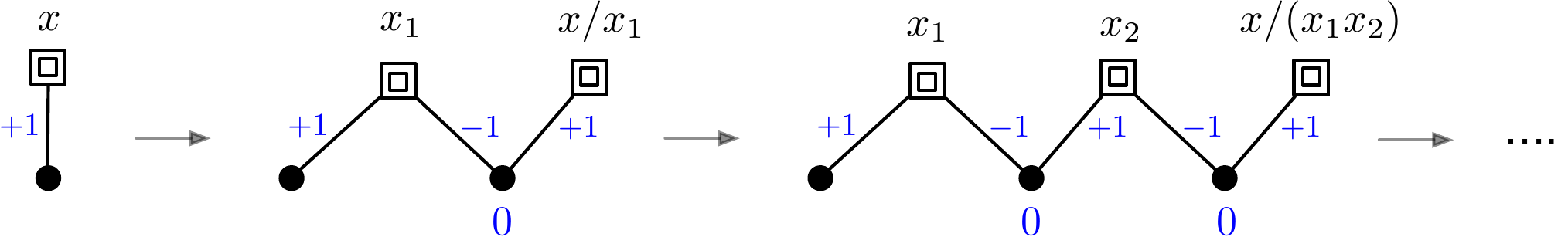}
    \end{aligned}
\end{equation}
In this process  new surgery circles $-\bullet_0-$ are introduced on each step:
 $$
\{L_1\}~\rightarrow ~\{L_1, L_2\} ~\rightarrow ~\{L_1, L_2, L_3\} ~\rightarrow ~\cdots   \,.
$$
The fusion chain gives
\begin{align}
(x,\q)_n = \sum_{n\ge k_1\ge k_2\cdots \ge k_p \ge 0} 
& \begin{bmatrix}
n \\k_1
\end{bmatrix}
\begin{bmatrix}
k_1 \\k_2
\end{bmatrix}
\cdots
\begin{bmatrix}
k_{p-1} \\k_p
\end{bmatrix} \cdot 
(x_1,\q)_{n-k_1} (x_2,\q)_{k_1-k_2} \cdots (x_p,\q)_{k_{p-1}-k_p}   
\nonumber \\
& \cdot x_1^{k_1}x_2^{k_2}\cdots x_p^{k_p}    \cdot  \Big(\frac{x}{x_1x_2 \cdots x_p}, \q\Big)_{k_p} 
\end{align}
This is a process of slightly separating many overlapped branes, or generating new branes from old ones, which is equivalent to the one brane case that there is only a matter $(x,\q)_n$ on it. The $\q$-binormal comes from the open strings stretching between two almost overlapped 5-branes. In \cite{Gukov:2016ac}, a similar recursion is applied on knot invariants.

\subsection{Fusion as the connected sum of matter circles}

In the last section, we discussed that the handle slides for gauge circles are linear recombinations of surgery circles $\{L_i\}$. Suppose we have some matter circles, then the question is how to perform handle slides for matter circles. We have not solved this problem, but in a special case we notice that the connected sum of matter circles can be interpreted as the fusion identity. 

To begin with, the connected sum for gauge circles is done through a special node $\bullet_0$ which has a vanishing framing number.  For instance, if we have two gauge nodes connecting to $\bullet_0$, then these two nodes merge together: $-\bullet_{k_1}-\bullet_0-\bullet_{k_2}-  ~=~  -\bullet_{k_1+k_2}-$.
One can ask what if we replace $\bullet$ by $\sqbox$ or $\dbox$? Could two matter nodes merge? The answer is roughly yes, but only precise identities for vortex partition functions could tell us some details and confirm the guess, and the fusion identity \eqref{fusion} is such an identity.
In the fusion identity, each $\q$-Pochhammer product denotes a matter node, and the subscript $k$ implies the presence of a gauge node $\bullet_0$. Since the fusion identity does not sum up the flux number $n$ we can ignore its associated gauge circle as it can be added/linked whenever we want. Then we can roughly use graphs to represent this fusion identity:
\begin{equation}\label{fusioncirc}
\begin{aligned}
\includegraphics[width=2.5in]{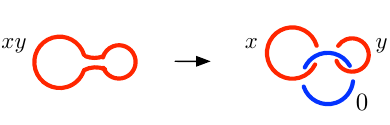}
    \end{aligned}
\end{equation}
where each red circle denotes a $\dbox$ \footnote{We apologize that in this note both $\dbox$ and $\sqbox$ are denoted by red circles $\redcirc$, and we hope to find a better notation in future to distinguish them. }. 
The fusion can be realized as a connected sum by the gauge node $\bullet_0$ (the blue circle with a vanishing framing number). 
Note that the matter circle $\redcirc(xy)$ could couple to other gauge nodes by gauging mass parameter.

\subsection{Fusion as Hanany-Witten transitions}
We can use the Hanany-Witten construction of 3d brane webs in \cite{Hanany:1996ie,Kitao:1999aa} to partially understand the fusion identity, although there are still a few subtle issues\footnote{I would like to thank Sung-Soo Kim and Futoshi Yagi for pointing out it.}.
Basically, the fusion identity can be interpreted as the Hanany-Witten move of a D5-branes and then the gauging of its flavor symmetry. We illustrate this process in the following brane web:
\begin{equation}\label{HWforfusion}
\begin{aligned}
\includegraphics[width=4in]{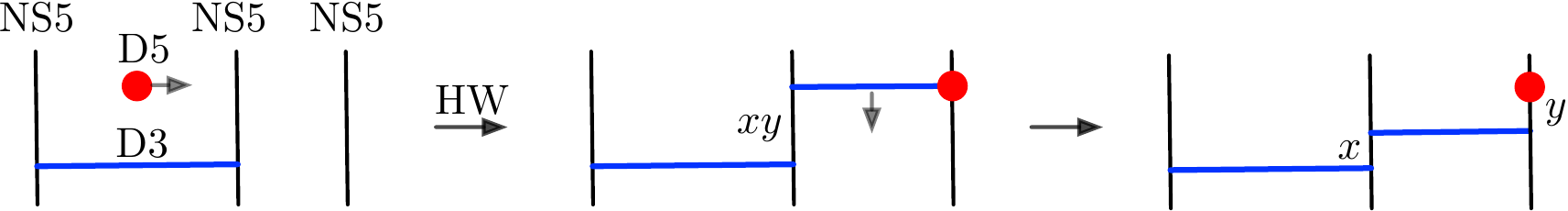}
    \end{aligned}
\end{equation}
where one additional NS5-brane is present, which gauges the flavor symmetry of the D3-brane when its attached D5-brane approaches this NS5-brane. One can then deform the position of the D3-brane to get the third web. One can continue this Hanany-Witten transition and gauging process to get a chain of webs for the fusion chain in \eqref{fusionchain}.

We need to remind that
 the FI parameters and mass parameters in the dual brane webs should be appropriately adjusted as in the fusion identity, such that their vortex partition functions could equal. The limit $y \rightarrow 0$ leads to the decoupling of a half of the D5-brane, and $y$ is the mass parameter for the anti-fundamental matter. The parameter $x$ is a free parameter for deforming a D3-brane if we keep the parameter $xy$ fixed.

The 3d brane webs could interpret many other 3d dualities. For instance,
\begin{equation} \label{fig:Sdualmatter}
\begin{aligned}
\includegraphics[width=3in]{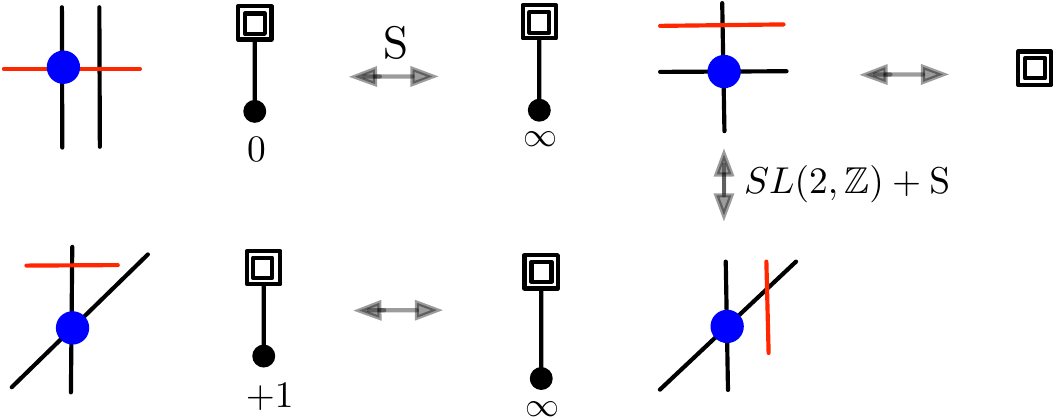}
    \end{aligned}
\end{equation}
where we draw the brane webs along a different direction of \eqref{HWforfusion}.
See e.g.\cite{Kitao:1999aa,Cheng:2023ocj,Cheng:2021vtq} for how the 3d brane webs encode Chen-Simons levels. In the above webs, the blue nodes denote D3-branes which are perpendicular to the plane of 5-branes. We use red lines to denote D5-branes to emphasize their positions under dualities.
In the first line, we draw the brane webs for the 3d $\N=4$ mirror pair $U(1)_0 + 1\dbox ~\leftrightarrow~ 1\dbox$. Since the node $\bullet_{\inf}$ can be deleted as we discussed before in section \ref{sec:STmove}, and the relative angle relates to the Chern-Simons level by $k=\tan\, \theta$, we can indirectly assign a brane web for a free hyper-multiplet $\dbox$.
In the second line, we show the brane webs for a pair of theories. Although the red lines in these two brane webs are related through a local $S$-duality in IIB string theory, it is not a duality. Because locally performing $S$-duality is mysterious, and the $S$-duality is usually applied on the whole brane web. From three-manifolds' perspective,
 red lines are  different 5-branes that come from the OV-defects wrapping on longitude or meridian of the surgery circle of $L(+1,1)$ respectively. In \cite{Cheng:2021vtq}, it is discussed that only when the anti-fundamantal matters for this pair are properly decoupled, the second line could be a duality, which is the special case that can be interpreted as equivalent twists of Lens spaces as $L(1,1)=L(1,n)$ for any $n \in \mathbb{Z}$.
After this proper decoupling, the special case is just  the brane webs for ST-duality $\bullet_{+1}-\sqbox ~\leftrightarrow~ \sqbox$, which 
 can also be obtained from the first line by decoupling. 

We can also perform handle slides on some brane webs. However, we have found a good correspondence in terms of brane webs. We only show some examples below:
\begin{equation}\label{threeHWex}
\begin{aligned}
\includegraphics[width=6in]{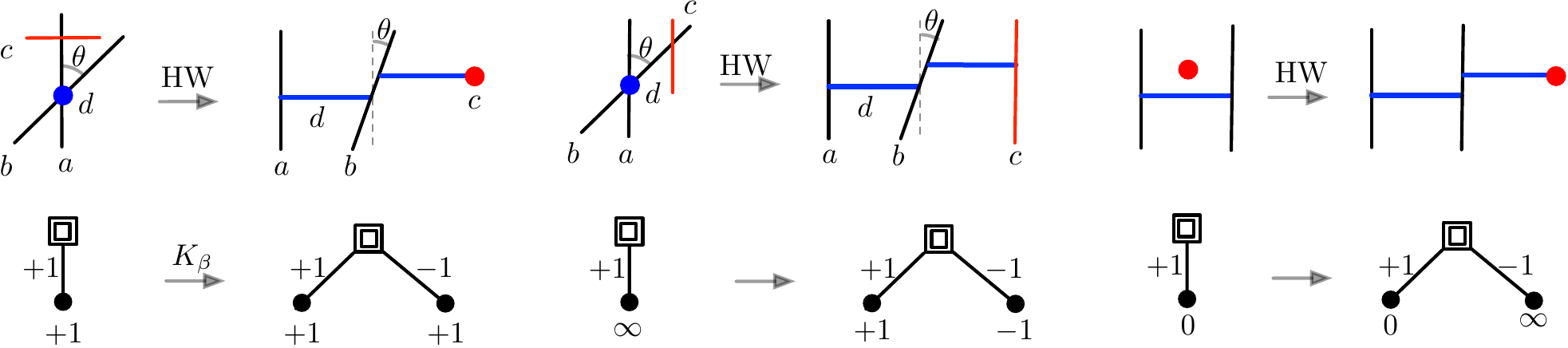}
    \end{aligned}
\end{equation}
For the last brane web, one can turn on CS level by rotating the first NS5-brane to get $\bullet_k-\dbox$. The above examples describe some descendent dualities of the fusion identity \eqref{fusion} by sending $y^{\pm1} \rightarrow 0$ and applying flips \eqref{variousflips}.

Moreover, we should be careful about the numbers of D3-branes in Hanany-Witten moves, since when a D5-branes crosses a $(p,q)$-brane, the number of introduced D3-brane is the intersection number: 
\begin{align}
\#(\text{D3})= \det \begin{bmatrix}
 1& 0 \\  p  & q 
\end{bmatrix} = q  \,.
\end{align}
In 3d brane web, we should set $q=1$ to match with the fusion identity.

\section{From fusion to its descendent dualities}

In this section, we find the fusion identity is highly non-trivial, as it can derive all dualities and their gauged versions that we know before and also that we have not known yet. 
													
\subsection{ST-duality}
To warm up, let us obtain the ST-dualities from the fusion identity by taking some decoupling limits, which are the settings of sending parameters $x$ or $y$ to $0$ or $\inf$.
For instance, if we send $ y\rightarrow0$, then the fusion identity reduces to
\begin{align}\label{fusionydep}
\frac{1}{(\q,\q)_n} = \sum_{k=0}^n \frac{ (x,\q)_{n-k} }{ (\q,\q)_{n-k} } \cdot \frac{x^k}{(\q,\q)_k} \,,
\end{align}
which is quite strange, since $x$ seems a  free mass parameter and emerges from nothing. We have not physically understand this identity, but we know that this identity reduces to \eqref{STreverse} in the limit $n \rightarrow \inf$, whose physical and geometric meaning is the $(ST)^2$-duality. 

Another case is that if we fix $xy=\q$, then fusion identity reduces to
\begin{align}
1 =
\sum_{k=0}^n \frac{ (x ,\q)_{n-k }}{ (\q,\q)_{n-k} } \cdot \frac{ (\q x^{-1},\q)_k}{(\q,\q)_k} \cdot x^k \,,
\end{align}
which is  the cancellation of two matter circles $\dbox$ through the connected sum. If we take the limit $n \rightarrow 0$, then this identity becomes
\begin{align}
\frac{ (\q,\q)_\inf} { (x,\q)_\inf } =\sum_{k=0}^\inf  \frac{ (\q x^{-1}, \q)_k}{ (\q,\q)_k} \cdot x^k  \,,
\end{align}
which is a special limit of the S-dual pair \eqref{inffusion}. From these two examples, one can image that the fusion identity could reduces to many other identities that may be interpreted as 3d dualities.

\subsection{Flipped S-dualtiy}\label{twistedST}
Let us derive a well known identity below
\begin{align}\label{fundtobi}
( x ,\q)_n = \sum_{j=0}^n 
\begin{bmatrix}
    n \\ j
\end{bmatrix}_q  
(-\sqrt{\q})^{j^2} ( x/\sqrt{\q})^j   \,,
\end{align}
which can be written as
\begin{align}\label{fundtobi2}
\frac{( x ,\q)_n}{(\q,\q)_n } = \sum_{j=0}^n
    \frac{1}{ (\q,\q)_{n-j} (\q,\q)_j } 
    \cdot   (-\sqrt{\q})^{j^2} ( x/\sqrt{\q})^j 
     \,.
\end{align}
At the large flux number limit $n \rightarrow \inf$, this identity reduces to the ST-duality \eqref{STduality}. Hence it seems that this identity describes a duality for theories with two chiral matters. 

Then the question is how to derive this identity and what causes this duality.
The short answer is that we can use fusion identity to derive it. Firsly, we define a twisted parameter $\tilde{x} = xy$ and use the flip identity in \eqref{variousflips} to rewrite 
$
 (y,\q)_k = (y^{-1},\q^{-1})_k (-\sqrt{\q})^{k^2} ( y/\sqrt{\q})^k 
$, and then the fusion identity \eqref{fusion} can be written as:
\begin{align}\label{flipdepy}
(\tilde{x},\q)_n = \sum_{k=0}^n
\bigg[
\begin{matrix} 
    n\\k
\end{matrix} \bigg]_\q
(\tilde{x} \cdot y^{-1},\q)_{n-k} (y^{-1},\q^{-1})_k \cdot 
(\tilde{x}/\sqrt{\q}  )^k (-\sqrt{\q})^{k^2}  \,.
\end{align}
If we vanish mass parameter by $y^{-1} \rightarrow 0$, then  two matters are decoupled and \eqref{fundtobi} is obtained. Recall that $y^{-1} \rightarrow 0$ is the large negative mass limit and $y\rightarrow 0$ is the large positive mass limit. One can observe that the two decoupling limits $y^{\pm1} \rightarrow 0$ lead to two very different identities or dualities in in \eqref{fusionydep} and \eqref{fundtobi} respectively.

We emphasize that actually \eqref{fundtobi} is the gauged version of the 3d $\mathcal{N}=4$ S-dual pair $U(1)_0 +1 \dbox~\leftrightarrow ~ 1 \dbox$ but its flipped version. The plumbing graphs for  \eqref{fundtobi}  and some equivalent versions given by handle slides and ST-moves are shown as follows
\begin{equation}
\label{Sdualhandle}
\begin{aligned}
\includegraphics[width=3.5in]{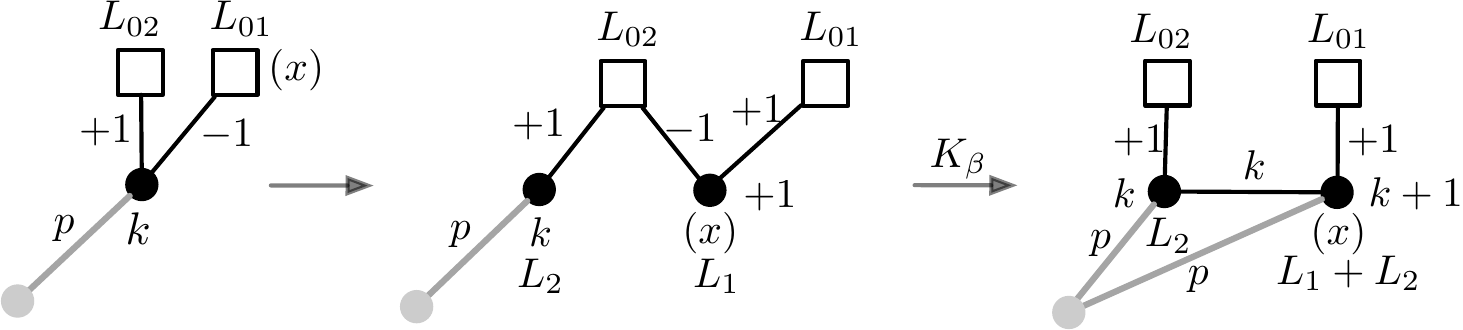}
    \end{aligned}
\end{equation}
Moreover, we clarify that the identity \eqref{fundtobi} is not a ST-move of an anti-fundamental matter up to handle slides, as the later's graphs in below are different from \eqref{Sdualhandle}:
\begin{equation}
\begin{aligned}\label{STmoveslidesex1}
\includegraphics[width=5.5in]{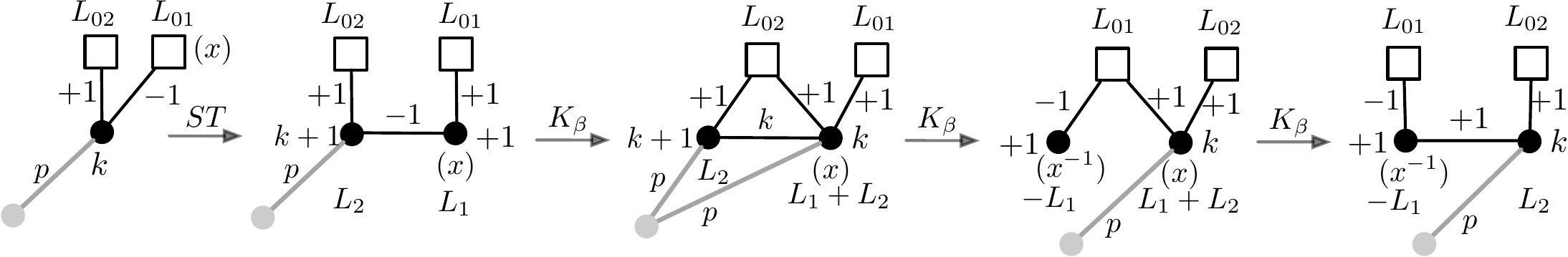}
    \end{aligned}
\end{equation}

Moreover,
in \cite{Cheng:2023ocj} we already derived this graphs \eqref{Sdualhandle} from sphere partition functions by partially gauging the SQED-XYZ duality. It is not a big surprise that this twisted/flipped S-duality can also derive \eqref{Sdualhandle}, as we will show later that SQED-XYZ duality arises from the fusion \eqref{fusion}.

\subsection{The S-duality and S-move }
In the above section, we show that a well know identity is the flipped S-dual pair, but have not clarified the S-dual pair \cite{Kapustin:1999ha,Benvenuti:2016wet}. In this section, we will fill this loophole. Basically, this S-duality borrows the name $\mathcal{S}$-duality in IIB string theory which exchanges D5-branes and NS5-branes and explains a 3d $\N=4$ mirror pair $ \bullet_0 -\dbox ~\leftrightarrow~\dbox $.

\vspace{2mm}
\noindent
\textbf{The limit $n=\inf$ and ungauging.} 
In the limit $n \rightarrow \inf$ or in other words the ungauging limit which removes the gauge circle linking to it, then  the fusion identity gives rise to
\begin{align}
\label{inffusion}
\boxed{
\frac{ (x y, \q)_\inf}{ (x,\q)_\inf } =\sum_{k=0}^\inf  \frac{ (y,\q)_k}{(\q,\q)_k} \cdot  x^k
}
\end{align}
which is already known to be an idenity for the vortex partition function of $U(1)_0 +1\F +1 \AF$, which is dual to a free hyper matter $1\F+1\AF$. Graphically, this duality is $ \bullet_0 -\dbox ~\leftrightarrow~\dbox $ that we mentioned.

There is an indirect brane web interpretation for this mirror duality pair in \cite{Cheng:2021vtq}. We show it below:
\begin{equation}\label{u1faf}
\begin{aligned}
\includegraphics[width=2.5in]{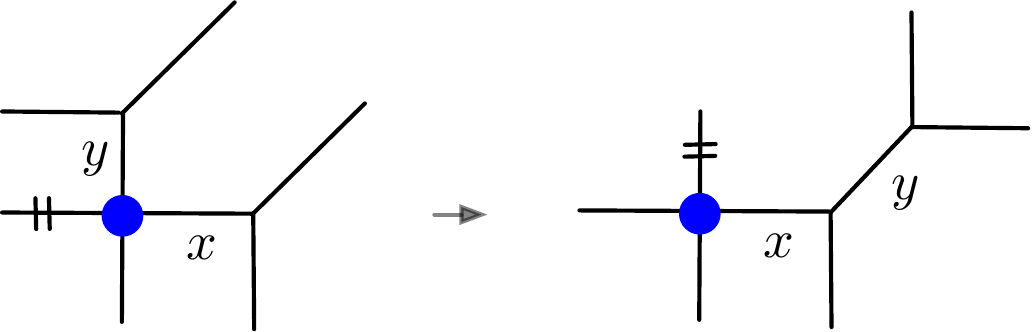}
    \end{aligned}
\end{equation}
which is interpreted as a Hanany-Witten transition of a 7-brane that is freely attached at the endpoints of the D5-branes at infinity, or the parallel movement of a half of D5-brane from one NS5-brane to the other NS5-brane that sandwiches a D3-brane. In \cite{Cheng:2021vtq}, topological vertex is used to compute the refined vortex partition functions. This computation formula needs to select preferred directions. For the webs in \eqref{u1faf}, the left web is given a preferred direction along horizontal lines, and for the right web the preferred direction is on vertical lines, as we have marked on the brane webs.

The right web describes the theory $\bullet_\inf - 1 \dbox$, for which $\bullet_\inf$ can be deleted to return a free hypermultiplet; see \eqref{fig:Sdualmatter} to get the right web up to a S-duality. The topological vertex computation can also show that the right web encodes a free hypermultiplet as the Young diagram on its preferred direction is empty, and the left web encodes a $U(1)_0+1 \dbox$. 
If we add one-loop parts, then the vortex partition functions for the left and right webs are:
\begin{align} \label{webSdualpart}
    Z^{U(1)_0 +1\F +1 \AF} = \frac{ (\q,\q)_\inf}{(y,\q)_\inf }  \cdot \sum_{k=0}^\inf  \frac{ (y,\q)_k}{(\q,\q)_k} \cdot x^k  =
   \frac{(\q, \q )_\inf }{ (y,\q)_\inf }  \cdot   \frac{(xy,\q)_\inf }{(x,\q)_\inf }  \,.
\end{align}
which is also \eqref{inffusion}.
Setting $x=\q$, this identity reduces to \eqref{xequqlq1}, so \eqref{inffusion} is a more generic case than \eqref{xequqlq1}.

Note that the left hand side of \eqref{webSdualpart} can be interpreted as gauging ${(z,\q)_\inf }/{(y,\q)_\inf} $ by shifting $ y \rightarrow y \q^k$ and $ z \rightarrow \q \cdot \q^{k}$ and $z$ is set to be massless. Finally, we sum up the flux number $n$ to reproduce the vortex partition function. This process is 
\begin{align}
\frac{ (z,\q)_\inf}{(y,\q)_\inf } ~\xrightarrow{\text{gauge} ~y, z}~\frac{(z,\q)_\inf}{(y,\q)_\inf} \cdot \frac{ (y,\q)_k}{(z,\q)_k }  
~\xrightarrow[\text{fix value} ~z=q]{\text{FI and CS gauge} ~x}~\frac{ (\q,\q)_\inf}{(y,\q)_\inf }  \cdot \sum_{k=0}^\inf  \frac{ (y,\q)_k}{(\q,\q)_k} x^k   ~\,.
\end{align}
Therefore, this S-dual pair is a gauging invariant theory $\mathcal{T}/\G = \mathcal{T}$ if we properly adjust various parameters.
Note that if we set $k=0$, then the gauging is turned off and we are only left with one-loop parts. This also implies that $k=\inf$ does not turn off the gauging but should be viewed as the counter terms of one-loop parts or some free chiral multiplets.

\vspace{2mm}
\noindent
\textbf{Gauge $x= x \q^{p n}$.} 
If gauging \eqref{inffusion}, this S-dual pair leads to a 2-2 move, which we call S-move. Firstly, we can gauge $x=x \q^{pn}$ in \eqref{inffusion} with a charge $p$ turned on to get
\begin{align}\label{Smoveide}
\boxed{
\frac{(xy,\q)_\infty}{(x,\q)_\inf} \frac{(x,\q)_{pn}}{ (xy,\q)_{pn}} =\sum_{k=0}^\inf \frac{(y,\q)_k }{(\q,\q)_k  } \cdot (-\sqrt{\q})^{2pnk} x^k
}
 \end{align}
which is the  S-move of a charge-$p$ hypermultiplet. There is two infinite $\q$-Pochhammer products in this formula which are the one-loop parts of the partition function. One can also add the one-loop parts to the chiral multiplets on the right hand side, and then correspondingly on the left hand we add $(\q,\q)_\inf/(y,\q)_\inf$. Then the factor is 
$  \frac{(xy,\q)_\inf (\q,\q)_\inf }{(x,\q)_\inf (y,\q)_\inf } $,
which is just the complete vortex partition function in \eqref{webSdualpart}.



 
One can read off mixed CS levels from this identity.
 The S-move could be represented by graphs:
\begin{equation}\label{fig:Smovefig}
\begin{aligned}
\includegraphics[width=1.3in]{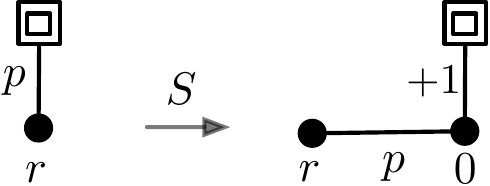}
    \end{aligned}
\end{equation}
 We can also flip the sign of the mass parameter $x$ using \eqref{variousflips} to get
 \begin{align}
\frac{(xy,\q)_\infty}{(x,\q)_\inf} \cdot
\frac{(\q x^{-1}y^{-1},\q)_{-pn}}{ (x^{-1},\q)_{-pn}} \cdot y^{-pn}=\sum_{k=0}^\inf \frac{(y,\q)_k }{(\q,\q)_k  } \cdot (-\sqrt{\q})^{2pnk} x^k  \,.
 \end{align}
 Therefore, in addition to \eqref{fig:Smovefig}, the S-move can be represented by graphs $\bullet_{k}-_{\pm p}\dbox ~\leftrightarrow~ \bullet_{k}-_{p} \bullet_0-\dbox$ where the linking number can have positive or negative signs depending on $p$.
 
  A useful limit is that when $y=\q$, the S-move \eqref{Smoveide} reduces to 
 \begin{align}
 \frac{(\q x,\q)_\inf}{(x,\q)_\inf} \cdot  \frac{(x,\q)_n}{(\q x,\q)_n} = \sum_{k=0}^\inf (\q^n x)^k  =\frac{1}{1-\q^n x} \,, 
 \end{align}
 which is a cancellation limit on the right hand side that $\dbox \rightarrow 1$.

\vspace{4mm}\noindent
\textbf{Gauging $y=y \q^{\rho m}$.} If further gauging $y=y \q^{\rho m}$, we have the fully gauged duality:
\begin{align}
\boxed{
\frac{(xy,\q)_\infty}{(x,\q)_\inf} \frac{ (x,\q)_{pn}}{ (xy,\q)_{\rho m+ pn}} =\sum_{k=0}^\inf \frac{(y,\q)_{\rho m+k} }{ (y,\q)_{\rho m} (\q,\q)_k  } \cdot (-\sqrt{\q})^{2pnk} x^k
}  \,.
 \end{align}
We can turn off the gauging of $x$ by setting $p=0$; then the above reduces to the gauging only for $y$:
\begin{align}\label{Sgaugey}
\boxed{
\frac{(xy,\q)_\infty}{(x,\q)_\inf} \frac{(y,\q)_{\rho m}}{ (xy,\q)_{\rho m}} =\sum_{k=0}^\inf \frac{(y,\q)_{\rho m+k} }{  (\q,\q)_k  } \cdot x^k
} \,.
 \end{align}
 Note that we cannot set $y \rightarrow 0$ for \eqref{Sgaugey}, as this leads to a conflict with the identity \eqref{ST2move}. Since gauging and decoupling are not two commutative operations, if we decouple a parameter, then it disappears and we cannot make it dynamic again by gauging.

If replacing $\rho m $ by $n$ and fixing the value of $y$, namely $y \rightarrow \q \cdot \q^{n}$, the above identity reduces to a well known identity
\begin{align}
\frac{1}{(x,\q)_{n+1}} =
\sum_{k=0}^\inf
\begin{bmatrix}
    n+k\\k
\end{bmatrix}_\q
x^k  \,.
\end{align} 
This identity means that if we sum up all massless open strings (the massless bi-fundamental with flux number $n+k$) between two almost overlapped D3-branes with flux numbers $n$ and $k$ respectively, then we can get a massive chiral multiplet.
 It is not obvious if this identity describes a basic physical duality.
By using 
\begin{align}
\frac{(\q x,\q)_\inf}{(x,\q)_\inf} =\frac{1}{1-x}\,, ~~~~\frac{(\q x, \q)_k}{(x,\q)_k} =\frac{1-x \q^k}{1-x} \xrightarrow{x=\q} \sum_{j=0}^k \q^j \,,
\end{align}
one can rewrite this identity as
\begin{align}
\frac{(\q^{n+1}x,\q)_\inf}{(\q^{n}x,\q)_\inf} \cdot \frac{1}{(x,\q)_{n}} =
\sum_{k=0}^\inf
\begin{bmatrix}
    n+k\\k
\end{bmatrix}_\q
x^k  \,.
\end{align}

\subsection{Massless limits for fusion identity}
When the mass parameters in the fusion identity \eqref{fusion} are massless, namely $x=\q$ or $y=\q$, the contribution of a pair of brane and anti-brane cancel, namely $\doubsquare \rightarrow 1$. 
In massless limits, the fusion identity reduces to some interesting identities.

\vspace{2mm}
\noindent
\textbf{The case $y=\q$.} 
In this case, matters encoded in the second matter circle cancel each other, and \eqref{fusion} reduces to a new equivalence:
\begin{align}\label{yqlimit}
 \frac{ (\q x, \q)_n }{ (\q, \q)_n } =
\sum_{k=0}^n \frac{ (x ,\q)_{n-k }}{ (\q,\q)_{n-k} } \,\cdot  x^k  \,,
\end{align}
which turns out to be the gauging of the flavor symmetry  of a hypermultiplet that is turned into a bi-fundamental hypermultiplet. This process is similar to the gauging in \eqref{guagingmatter}. We can use the reduced graph of \eqref{fusionchain} to present this case:
\begin{equation}\begin{aligned}
\includegraphics[width=2in]{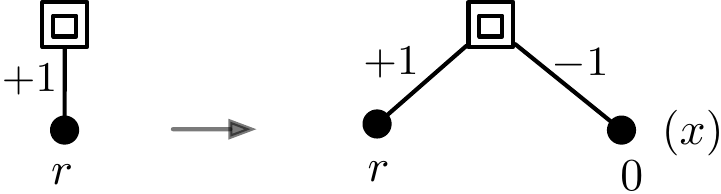}
    \end{aligned}
\end{equation}
This identity and the associated plumbing graphs can be interpreted through brane webs by adding one additional NS5-brane and Hanany-Witten moves as in \eqref{HWforfusion}. 
Note that this identity does describe a handle slides of a hypermultiplet, analogous to the slide in \eqref{slidezerofund}. What is interesting is that by adjusting FI parameter to equal to the mass parameter $x$, this theory is invariant under this handle slide. 

If we do a handle slide on the right hand side of \eqref{yqlimit} by shifting $k \rightarrow n-k$, an equivalent expression can be obtained:
\begin{align}
\boxed{
 \frac{ (\q x, \q)_n }{ (\q, \q)_n } \cdot  x^{-n}=
\sum_{k=0}^n \frac{ (x ,\q)_{k }}{ (\q,\q)_{k} } \, \cdot x^{-k}   }
\,,
\end{align}
for which the surgery circles on the right hand side change from $\{L_1, L_2 \}$ to $\{L_1+L_2 , -L_2  \}$, where $L_1$ is associated with the node $\bullet_{[n]}$ and $L_2$ is with $\bullet_{[k]}$ and $x$ is a FI parameter.

\vspace{2mm}
\noindent
\textbf{The case $x=\q$.} 
In this case the fusion identity reduces to
\begin{align}
\boxed{
 \frac{ (\q y, \q)_n }{ (\q, \q)_n } =
\sum_{k=0}^n \frac{ (y ,\q)_{k }}{ (\q,\q)_{k} } \, \cdot \q^k     }   \,.  
\end{align}
If we shift $k \rightarrow n-k$, then it becomes
\begin{align}
\frac{ (\q y, \q)_n }{ (\q, \q)_n }  =
\sum_{k=0}^n \frac{ (y ,\q)_{n-k }}{ (\q,\q)_{n-k} } \, \cdot \q^{n-k} \,. 
\end{align}
Similarly, if further taking $n=\inf$, 
one can get
\begin{align}\label{xequqlq1}
 \frac{ (\q y, \q)_\inf }{ (\q, \q)_\inf } =
\sum_{k=0}^\inf \frac{ (y ,\q)_{k }}{ (\q,\q)_{k} } \cdot \q^k    \,, 
\end{align}
which is a special case that $x=\q$ for \eqref{inffusion}. If viewing $\q y$ as the K\"ahler parameter for the resolved conifold, then conifold transition tells that this is equivalent to introduce a new D3-brane between D5 and NS5.

\section{Superpotentials arise from fusions}\label{sec:triangles}

The geometric interpretation for dualities involve superpotentials remains a problem. We have not clearly witnessed a superpotential for ST-moves or S-moves  in the above sections. In \cite{Cheng:2023ocj}, the author derived the gauged version of the SQED-XYZ duality, in which the duality involves a cubic superpotential. In this section, we derive this duality from the fusion identity and hence the superpotential is given an indirect geometric realization as fusions of matter circles.

\subsection{Superpotential triangles}

Superpotentials in 3d abelian theories are polynomials of chiral multiplets. The most basic superpotential is a cubic term in the XYZ model, which couples three matter fields and takes the form $\W =XYZ$.  The QED is the theory $U(1)_0 + 1\F+1\AF$. This mirror duality can be understand as the 3d $\N=4$ pair $U(1)_0 +1 \F +1\AF +1 \textbf{Adj}$ which is dual to a free hypermultiplet $1\F+1\AF$. Recall that the former has a cubic superpotential that couples three matter fields.

The gauged SQED-XYZ duality is obtained by gauging the global symmetries on both sides, which is $U(1) \times U(1)$. In the SQED, since the mass parameters for two chiral multiplets are equivalent, so the flavor symmetry is $U(1)_F$ and the other $U(1)$ is the topological symmetry $U(1)_T$ associated to the gauge group. In the XYZ-model, since it contains a cubic superpotential, both $U(1)$s are flavor symmetries. What is interesting is that the gauging of the global symmetries does not change superpotentials.

It turns out that gauged SQED-XYZ duality is a duality between two nodes graphs and three nodes graphs, and hence defines a two-three move. The three nodes graphs have three equivalent cases that are called cases $\textbf{A}, \textbf{B} , \textbf{E}_{r,s}$ or unlinking, linking, and exotic as one likes. We copy the results from \cite{Cheng:2023ocj} in below:
\begin{equation}\label{triangle}
\begin{aligned}
\includegraphics[width=5in]{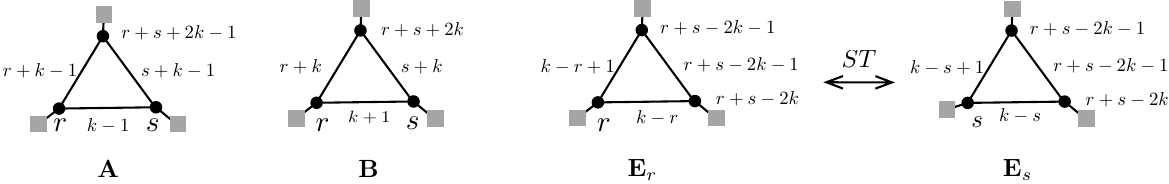}
    \end{aligned}
\end{equation}
which encode the same cubic superpotential $\W=XYZ$. One can observe that there are relations between the Chern-Simons levels or in other words the linking numbers, and these relations reflect the effect of superpotentials. 

The question is how to geometrically understand these triangles and what are the geometric structures that cause the SEQD-XYZ duality. The short answer is that the fusion of matter circles \eqref{fusion} is almost the operation of adding a cubic superpotential or in other words performing a Seiberg-duality in the sense that the SEQD-XYZ duality  is a Seiberg-duality.

Before providing details to this answer. We firstly simplify this 2-3 move by applying handle slides.
The unlinking process can be simplified as follows
\begin{equation} \label{unlinking}
\begin{aligned}
\includegraphics[width=5in]{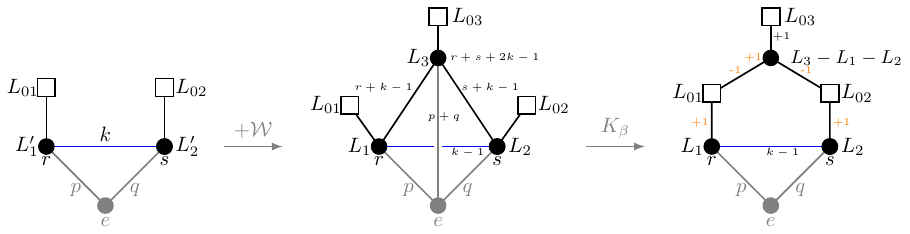}
    \end{aligned}
\end{equation}
where the node $\bullet_e$ is an external gauge node which also interacts with the 2-3 move as shown in the first step of \eqref{unlinking}. Note that the 2-3 move changes the surgery circles from $L_i' $ to $L_i$, which are not precisely the same. The additional gauge node $L_3$ decorated with a matter node $L_{03}$ is introduced because of the superpotential. 

The second step is a handle slide and we have assigned the associated charges and recombinations of surgery/gauge circles on the graphs. Interestingly, the handle slides could remove the interactions between the external node $\bullet_e$ and the addition node $L_3$, and many other links in the triangle are removed. We are left with a very simple graph that three matter nodes are coupled to a common gauge node as shown in the third graph. Recall that handle slides on gauge nodes do not affect matter nodes themselves but only change their charges. One can observe that the combined operation $K_\b  \circ \mathcal{W}$ turns fundamental matters $L_{01}$ and $L_{02}$ into bifundamental matters and one addition matter $L_{03}$ emerges. 
The linking number is slightly reduced from $L_1' \circ L_2'=k$ to $L_1 \circ L_2 =k-1$.

Moreover, the superpotential should exist at the node $L_3-L_1-L_2$ at where three matters are interacting. This can be understood from the brane configuration D3-NS5-D3 on which if one puts a D5 on NS5, then a cubic superpotential appears at the intersection of these three types of branes, and the interacting open strings stretch between D3-D3 and D3-D5, which give two bi-fundamental matters and one fundamental matter. This brane configuration is analogous to the third graph in  \eqref{unlinking}.

The other cases of superpotential triangles including linking case and exotic case can be simplified as well. In the linking case, we get the following transformations:
\begin{equation}\label{linkingform}
\begin{aligned}
\includegraphics[width=5in]{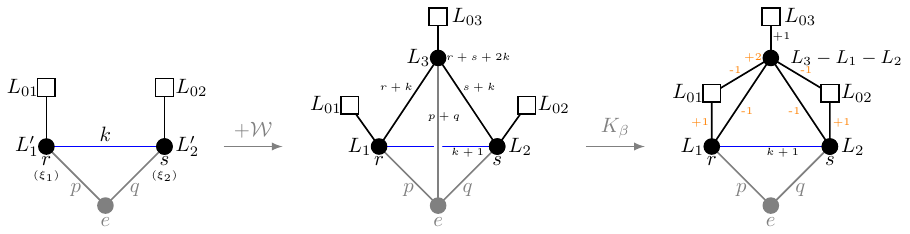}
\end{aligned}
\end{equation}
where one can see from the linking numbers that $L_3=L_1'+L_2'$, and there is an almost equivalence $L_{1,2}\sim L_{1,2}'$ but only differ by the linking number $L_1\cdot L_2 \neq L_1' \cdot L_2'$.

For the exotic cases, we get the following:
\begin{equation}\label{exoticcase}
\begin{aligned}
\includegraphics[width=5in]{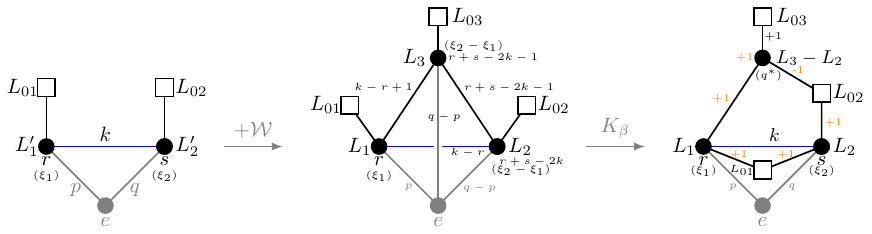}
\end{aligned}
\end{equation}
where we have assigned FI parameters $\xi_i$ on graphs. One can observe that $L_{1,2}'$ relate to $L_{1,2}$ by the relations $L_1=L_1'\,, ~L_2=L_2'-L_1'$, and hence $L_2+L_1=L_2'$. The exotic case $\textbf{E}_s$ is just equivalent to the exotic case $\textbf{E}_r$ by switching $r$ and $s$.

\subsection{Fusion to unlinking}
In \cite{Ekholm:2019lmb} and in particular the answer of Fedor Petrov on the website “mathematics stack exchange”, the third graph in \eqref{unlinking} had been partially derived in a combinatorial way, in which they derive the identity 
\begin{align}\label{nongenunlink}
\frac{\q^{ab}}{ (\q,\q)_a (\q,\q)_b } = \sum_{k=0}^{ \text{min}(a,b)} 
\frac{ (\sqrt{\q})^{k^2-k} (-1)^k }
{ (\q,\q)_{a-k} (\q,\q)_{b-k} (\q,\q)_k } \,.
\end{align}

We claim that the application of  gauging, decoupling, and flips  on fusion identity provides a more complete derivation. Basically, we perform the flipped gauging $x \mapsto x \q^{-a}$ and decouple $y \mapsto 0$.
Before doing that, we perform a handle slide $k \rightarrow b-k$ on the fusion identity to get
\begin{align}
\frac{(xy,\q)_b}{(\q,\q)_b} \cdot x^{-b} = \sum_{k=0}^b  \frac{(y,\q)_{b-k}}{(\q,\q)_{b-k}} \cdot  \frac{(x , \q )_k}{(\q,\q)_k} \cdot x^{-k} \,.
\end{align} 
If the gauging $x \mapsto x \q^{-a}$ is applied, we get
\begin{align}\label{flipgaugefusion}
\frac{(xy \q^{-a},\q)_b}{(\q,\q)_b} \cdot x^{-b} \q^{ab} = \sum_{k=0}^b  \frac{(y,\q)_{b-k}}{(\q,\q)_{b-k}} \cdot  \frac{(x \q^{-a} , \q )_k}{(\q,\q)_k} \cdot x^{-k} \q^{ak} \,.
\end{align}
We can roughly draw the following graph for this identity 
\begin{equation}
\begin{aligned}
\includegraphics[width=3in]{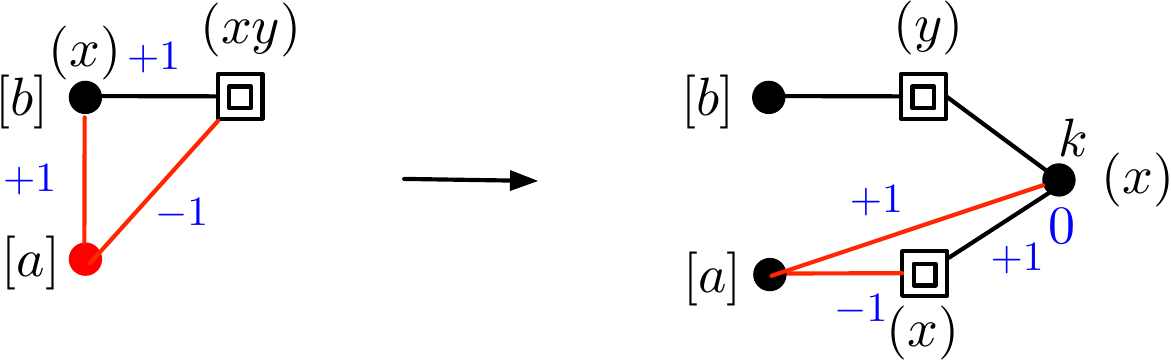}
    \end{aligned}
\end{equation}
where the red lines are the links caused by gauging, and the red node is for the introduced gauge node $\bullet_{[a]}$ labeled by flux number $a$. We emphasize various parts of the graphs by colors. Note that although in the hypermultiplet $\dbox$ only the anti-fundamental matter is massive, we can still roughly draw the gauged graph and assign the charges for the $\dbox$.

Now we are going to take into the below formula
\begin{align}
\boxed{
{(x  \q^{-a},\q)_k}=
 \frac{ (\q x^{-1},\q)_a}{ (\q x^{-1},\q)_{a-k}}  \cdot
(-\sqrt{\q})^{k^2-2ak}  \left(\frac{x}{\sqrt{\q}}\right)^k  }
 \,,
\end{align}
which is basically flipped gauging and can be derived by combining \eqref{variousflips} and \eqref{gaugemass12} \footnote{In this case, $(x,\q)_{a-k} =(x,\q)_a( \q^a x,\q)_{-k}$}.
After decoupling a chiral multiplet by setting $y=0$, some terms cancels and decouples, and \eqref{flipgaugefusion} becomes 
\begin{align} \label{genericunlink}
\boxed{
\frac{ (-\sqrt{\q})^{2a b}  x^{-b} }{ (\q x^{-1},\q)_a (\q,\q)_b } =
\sum_{k=0}^{b }
\frac{ (-\sqrt{\q})^{k^2}  (1/ \sqrt{\q})^{k}   }{ (\q x^{-1},\q)_{a-k} (\q,\q)_{b-k}   (\q,\q)_k  }
} \,.
\end{align}
The identity describe the linking case of superpotential triangles in \eqref{unlinking}. The background linking numbers $r$, $k$ and $s$ can be turned on by hand. One can see that this form is more generic than \eqref{nongenunlink}, as it shows that one mass parameter $x$ can be turned on.
Finally, the corresponding graph is
\begin{equation}
\begin{aligned}
\includegraphics[width=3in]{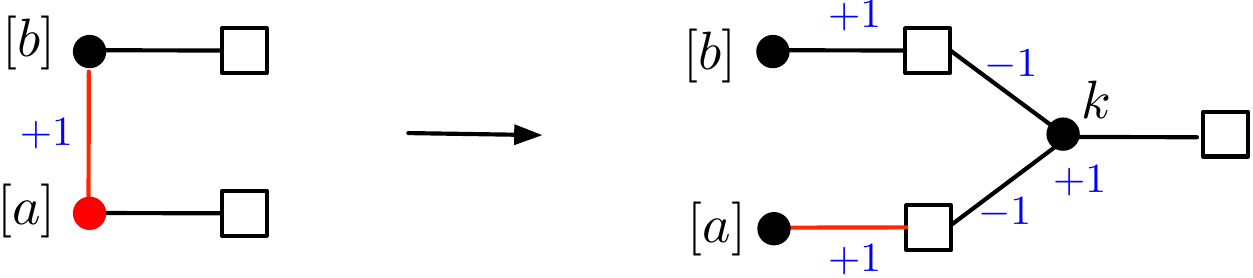}
    \end{aligned}
\end{equation}
which is just the third graph in \eqref{unlinking}.
Moreover, the gauging also tells us that we can also choose other charges $p$ by shift $a \mapsto pa$, which naively changes the formula.

We want to emphasize the phenomenon that
a bifundamental hypermultiplet can be written in terms of chiral multiplets:
\begin{equation}
\begin{aligned}
\includegraphics[width=3in]{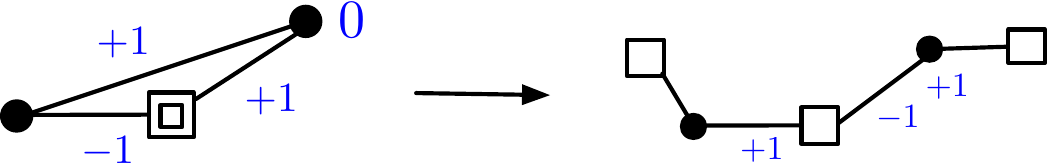}
    \end{aligned}
\end{equation}
In terms of $\q$-Pochhammer product, this is the process that 
\begin{align}
\frac{(x,\q)_k}{(\q,\q)_k} ~\xrightarrow{x \rightarrow x \q^{-a}} ~
\frac{(x  \q^{-a},\q)_k}{ (\q,\q)_k} =
 \frac{ (\q x^{-1},\q)_a}{(\q,\q)_k (\q x^{-1},\q)_{a-k}}  \cdot
(-\sqrt{\q})^{k^2-2ak}  \left(\frac{x}{\sqrt{\q}}\right)^k \,,
\end{align}
which in the case that $x=1$ describes the almost overlapped D3-branes.

\vspace{4mm}\noindent
\textbf{More descendent identities (dualities).}
By using the move \eqref{STreverse}, \eqref{ST2move} and turning on a FI-term $z^b$ on both sides of unlinking \eqref{genericunlink} and summing up $b$, namely $\sum_{b=0}^{\inf} z^b $,  one can get a formula similar to  \eqref{fundtobi2} as follows\footnote{On the right hand side, we did the handle slide $b \rightarrow b+k$ for the sum.}
\begin{align}\label{sdualgauge}
\boxed{
\frac{1}{(z x,\q)_\inf}\cdot \frac{ (z x,\q)_a}{(\q x,\q)_a} =\frac{1}{(z,\q)}_\inf \cdot
\sum_{k=0}^{a}\frac{(-\sqrt{\q})^{k^2} (z/\sqrt{\q})^k }{ (\q x,\q)_{a-k}   (\q,\q)_k }    }  \,,
\end{align}
where we have replaced $x \rightarrow x^{-1}$ to let it looks nicely. This identity reduces to \eqref{fundtobi2} at $x=1$ and hence is a generic version of the corresponding flipped S-dual pair. 

Similarly, if we sum up $b$  and turn on a particular CS level $\sum_{b=0}^{\inf} z^b (-\sqrt{\q})^{b^2-b}$ and replace $1/x \rightarrow x$ for the unlinking \eqref{genericunlink}, then we have
\begin{align}\label{partguage2}
\boxed{
\frac{ (z x, \q )_\inf  }{ (z x, \q )_a   (\q x, \q )_a } = { (z, \q )_\inf } \cdot \sum_{k=0}^\inf \frac{  (-\sqrt{\q})^{2 k^2} (z/\q)^k      }{ (\q x,\q)_{a-k}  (\q, \q )_k  (z,\q)_k       }    } \,.
\end{align}
The above two new identities \eqref{sdualgauge} and \eqref{partguage2} describe the two cases of partial gauging, namely gauging only a $U(1) \subset U(1) \times U(1)$. In \cite{Cheng:2023ocj} this partial gauging was done by sphere partition functions. Since the SQED-XYZ duality is a self-gauging duality, namely $\mathcal{T}/\mathcal{G}=\mathcal{T}$, integrating out the gauged parameters leads to itself and hence implies ungauging. This explains why the partial gauging emerges from the fully gauged case.

An interesting limit for the above two identities is given by setting $x=1$, then we have respectively
\begin{align}
(z, \q)_a &=\sum_{k=0}^\inf \begin{bmatrix}a\\ k  \end{bmatrix}_\q
(-\sqrt{\q})^{k^2} ( z/ \sqrt{\q})^k  \,,\\
\frac{1}{ (z,\q)_a }&= \sum_{k=0}^\inf  \begin{bmatrix}a\\ k  \end{bmatrix}_\q  \frac{ (-\sqrt{\q})^{2 k^2} (z/\q)^k  }{ (z,\q)_k  } \,,
\end{align}
where the first identity is the flipped S-duality \eqref{fundtobi2}. These two identities seem dual to each other. However, we have not recognized the second identity as a duality. 

One can take the limit $a=0$ of \eqref{genericunlink} to obtain the identity
\begin{align} 
\frac{ x^{-b} }{  (\q,\q)_b } =
\sum_{k=0}^{b }
\frac{ (-\sqrt{\q})^{k^2}  (-1)^{k}   (x,\q)_k}{  (\q,\q)_{b-k}   (\q,\q)_k  }   \,,
\end{align}
which is a $1-3$ move. 
We can also take some other limits like $a \rightarrow 0, \inf$ or $\b \rightarrow 0, \inf$ for the linking, unlinking and exotic cases that we will discuss later. These limits may lead to more versions of $1-3$ moves.


\subsection{Fusion to linking and exotic}
Once the crucial role played by gauging is noticed, it becomes easy to derive other cases of superpotential triangles from fusion.

To begin with, let us rewrite \eqref{flipgaugefusion} as
\begin{align}
x^{-b} \q^{a b}  \cdot \frac{ ( xy \q^{-a},\q)_b}{ (\q,\q)_b }  =
\sum_{k=0}^b 
\frac{ ( y,\q)_{b-k} (\q x^{-1} ,\q)_a }{ ( \q,\q)_{b-k}  (\q x^{-1} ,\q )_{a-k} (\q,\q)_k }  \cdot 
(-\sqrt{\q})^{k^2} (1/\sqrt{\q})^{k}  \,.
\end{align}
Flipping all mass parameters on the left hand side gives
\begin{align} 
\label{gaugexkeepy}
\frac{  (\q x^{-1} y^{-1} ,\q  )_a }{ (\q x^{-1} y^{-1} , \q )_{a-b} (\q,\q)_b } \cdot (-\sqrt{\q})^{b^2} (y/\sqrt{\q})^b 
=
 \sum_{k=0}^b 
\frac{ ( y,\q)_{b-k} (\q x^{-1} ,\q)_a }{ (\q,\q)_{b-k} (\q x^{-1} ,\q )_{a-k} (\q,\q)_k } \cdot 
(-\sqrt{\q})^{k^2} (1/\sqrt{\q})^{k}  \,.
\end{align}
In this case, we can decoupe $x^{-1}\rightarrow 0$ to get
\begin{align}
 (-\sqrt{\q})^{b^2} (y/\sqrt{\q})^b 
=
 \sum_{k=0}^b \begin{bmatrix}
b\\k
\end{bmatrix}_q \cdot
 ( y,\q)_{b-k}  \cdot 
(-\sqrt{\q})^{k^2} (1/\sqrt{\q})^{k}  \,.
\end{align}
After performing the handle slide $k \rightarrow -k +b$ and gauging $y=y \q^{-a}$ on this identity, we derive the linking form for \eqref{linkingform}:
\begin{align}\label{linkingfromfusion}
\boxed{
\frac{  (-\sqrt{\q} )^{-2 a b}  y^b}{ (\q y^{-1}, \q)_a (\q,\q)_b }  = \sum_{k=0}^b  \frac{ ( -\sqrt{\q})^{ 2k^2-2ak-2bk } y^k }{ (\q y^{-1}, \q)_{a-k}  ( \q,\q)_{b-k}  (\q,\q)_k }
} \,.
\end{align}
One can check that the unlinking formula \eqref{genericunlink} relates to the linking formula \eqref{linkingfromfusion} by the orientation flip $\q \rightarrow \q^{-1}$ and replacing $x \rightarrow y^{-1}$ and up to a shift on FI parameter $(\sqrt{\q})^{a+b}x^{-a}$, so unlinking and linking are roughly equivalent by flipping the orientation of the three-manifold.

To derive exotic superpotential triangles, one needs to apply flipped gauging $y =y \q^{-a}$ to get 
\begin{align}
\label{gaugeykeepx}
\frac{ ( \q x^{-1} y^{-1}, \q)_a}{ (\q x^{-1}y^{-1}, \q  )_{a-b}}  =
\sum_{k=0}^b \begin{bmatrix}
b\\k
\end{bmatrix}_q \cdot
 \frac{( x,\q)_{k} (\q y^{-1}, \q )_a} {  (\q y^{-1} ,\q )_{a-b+k} }
  \cdot 
(-\sqrt{\q})^{k^2+2 a k -2 b k} ( \sqrt{\q} x^{-1}y^{-1})^{k} \,.
\end{align}
This identity is almost the same as \eqref{gaugexkeepy} if switching $x$ and $y$, except that the factor on the right hand side of \eqref{gaugexkeepy} can be reorganized as $ (-\sqrt{\q})^{k^2 - 2 b k} (\sqrt{\q})^k y^{-b}$, while that of \eqref{gaugeykeepx} is $(-\sqrt{\q})^{ k^2+2 a k -2 b k} ( \sqrt{\q} x^{-1}y^{-1})^{k}$.
In the case \eqref{linkingfromfusion}, we cannot set $ x^{-1}$ or $y^{-1}$ to be zero for decoupling, as this vanishes the whole identity.
The identity \eqref{gaugeykeepx} can reduce to the superpotential triangles in \eqref{exoticcase} by setting $y^{-1}=0$ but fixing $\tx:=x^{-1}y^{-1}$, and a handle slide $a \rightarrow a+b$ to get the following form for exotic triangle:
\begin{align}\label{fusiontoexotic}
\boxed{
\frac{1}{(\q \tx,\q )_a (\q,\q)_b }  = \sum_{k=0}^b \frac{ (-\sqrt{\q})^{k^2+2 a k} (\sqrt{\q} \tx)^k }{  (\q \tx ,\q)_{a+b} (\q,\q)_{b-k} (\q,\q)_k  }  
} \,.
\end{align}
By taking the limit $b\rightarrow \inf$, the above formula reduces to the ST-move in \eqref{STmovequiver}.
We can also take $a \rightarrow 0$ to get \eqref{fundtobi2}.

\section{Webs of dualities}\label{sec:structures}

One can draw a big diagram to represent the connections between various dualities that arise from the fusion identity, although by now we have not found a physical duality for describing the fusion identity, many of its descendent identities are physical dualities. 
From the fusion mother, we can draw the duality webs:
\begin{equation}
\label{fusiontoW}
\begin{aligned}
\begin{tikzpicture}
\node at (0,0) {~~$\text{fusion}\, \eqref{fusion}$} ;
\draw[->, thick] (-0.5,-0.5)--(-1.9,-1.5)  node[midway,left=0.2]{$ \G_-(x)$};
\draw[->, thick] (-2.2,-1.8)--(-4,-3) node[midway,left]{$y\rightarrow0~$} node[below]{$\text{unlinking} \, \eqref{genericunlink}$};
\draw[->, thick] (-0,-0.5)--(0,-1.5)  node[midway,right]{$ \G_-(y)$};
\draw[->, thick] (0,-1.8)--(0,-3) node[midway,right]{$x^{-1}\rightarrow0~$} node[below]{$\text{linking} \, \eqref{linkingfromfusion}$};
\draw[->, thick] (0.5,-0.5)--(2.2,-1.5) node[midway,right=0.4]{$ \G_-(x^{-1}y^{-1})$};
\draw[->, thick] (2.5,-1.8)--(4.3,-3) node[midway,right]{$~~y^{-1}\rightarrow0~$} node[below]{$\text{exotic} \, \eqref{fusiontoexotic}$};
;
\end{tikzpicture}
\end{aligned}
\end{equation}
\begin{equation}\label{fusiontoST}
\begin{aligned}
\begin{tikzpicture}
\node at (0,0) {$~~\text{fusion}\, \eqref{fusion}$} ;
\draw[->, thick] (-0.5,-0.5)--(-1.9,-1.5)  node[midway,left=0.2]{$n \rightarrow \inf$}node[below]{$\text{S-duality} \, \eqref{inffusion}$};;
\draw[->, thick] (-2.8,-2.2)--(-4.5,-3.5) node[midway,left]{$\G_+(x)~~$} node[below]{$\text{S-move} \, \eqref{Smoveide}$};
\draw[->, thick] (0.5,-0.5)--(4.3,-3)node[below]{$\text{flipped S-duality} \, \eqref{fundtobi2}$}  ;
\node at ( 2.2,-1 ){~$\text{fix} ~xy$};
\node at (3.8,-1.6){$\text{and }y^{-1}\rightarrow0~$} ;
\draw[->,thick](4.3,-3.7)--(4.3,-4.7) node[right,midway]{$n \rightarrow \inf$}node[below]{$\text{ST-duality}~\eqref{STduality}$};
\draw[->,thick](4.3,-5.4)--(4.3,-6.2) node[right,midway]{$\G_+(x)$}node[below]{~~\quad$\text{ST-move}~\eqref{STmovequiver}$};
\draw[->,thick](3.2,-5.4)--(2,-6.2) node[left,midway]{$\G_-(x)$~~};
\node at (0.6,-6.6){$\text{eff. ST-move (Kirby)}~\eqref{STflipmass}$};
\draw[<-,thick] (2,-7)--(2,-8)node[below]{$\text{exotic} \, \eqref{fusiontoexotic}$} node[midway,left]{$b\rightarrow \inf$}  ;
\draw[<-,thick] (4.2,-7)--(2.8,-8) node[midway,right] {~$a\rightarrow 0$} ;
\draw[->, thick](-4.4,-4.2)--(-4.4,-5)--(2.5,-5) node[midway,above]{$\text{fix~} xy ~\text{and then}~ y^{-1}\rightarrow0$} ;
\end{tikzpicture}
\end{aligned}
\end{equation}
The symbol $\G_+(x)$ denotes the gauging of the parameter $x \rightarrow x \q^n$ which preserves the orientation of the parameter $x$, while $\G_-(x)$ means the flipped gauging $x \rightarrow x \q^{-n}$ which requires to flip the orientation of the parameter.

The skew rectangle in the bottom of this diagram shows a closed relation between the exotic case of superpotential triangles and the ST-moves that can be interpreted as the Kirby moves with maters.  Note that the above duality webs do not exhaust all relations; for instance, the twisted S-duality in \eqref{sdualgauge} can be obtained by summing up a gauge degree of the unlinking form \eqref{genericunlink}.

Since both fusion, gauging and flips have geometric interpretations, all descendent identities or dualities can inherit geometric interpretations from fusion and other operations.

The above diagrams shows some algebraic structures of 3d dualities, which we expect to be a modified Hopf algebra of quantum groups. In particular, the flip of mass parameters should be regarded as the antipode operator $S$ in the Hopf algebra, and the fusion identity  should be viewed as one crucial property of Hopf algebra. More study deserves to be done for this hiding algebraic structure.

\section{TFT structures}\label{sec:Witten}

In the above sections, we discuss that the fusion identity and 3d dualities enjoy the geometric structures of three-manifolds, which are also the structures of topological field theories on three-manifolds if taking into account 3d-3d correspondence.

In the 3d theories, the Ooguri-Vafa defects give chiral muletiplets, and these defects lead to Wilson loops of the 3d complex Chern-Simons theory on the same three-manifold. In principle, we should sum up the representations of the Wilson loops to get a topological string partition function. But if we ignore the summation, then a $\q$-Pochhammer product $(x,\q)_n/(\q,\q)_n$ corresponds to a Wilson loop with a winding number $n$. Therefore, we can regard matter circles as Wilson loops in \cite{witten89} and use topological field theory to analyze. 


\subsection{Surgeries and Wilson loops for the connected sum}

Let us go back to surgery to interpret the connected sum, which is through
a handle slide below
\begin{equation}
\begin{aligned}
\includegraphics[width=3.2in]{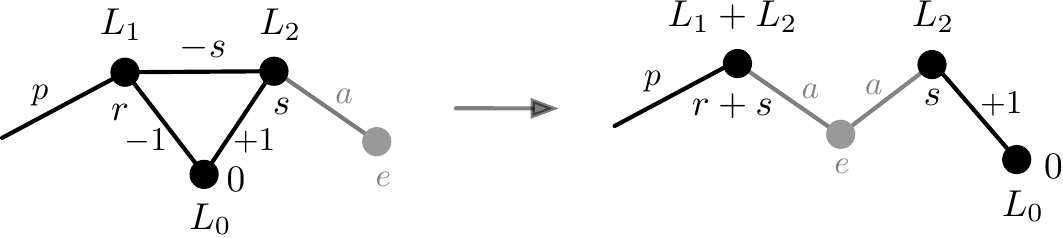}
    \end{aligned}
\end{equation}
where $\bullet_e$ is an external node. After the handle slide, this external node could join the graph, so it is more better to set $a=0$ and hence no external node is allowed.
After removing any external node, we have 
\begin{equation}
\begin{aligned}
\includegraphics[width=3.2in]{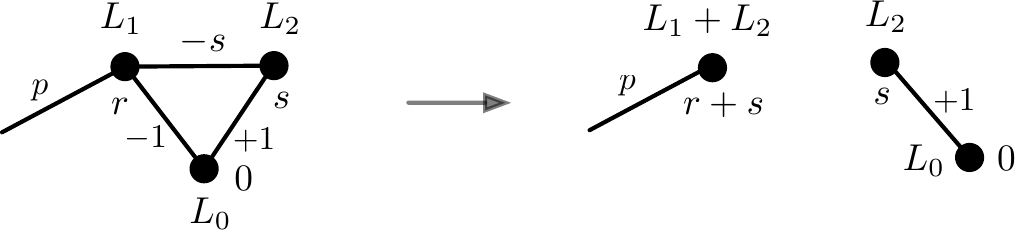}
    \end{aligned}
\end{equation}
To match with the fusion identity \eqref{fusion}, we should put Ooguri-Vafa defects to replace $\bullet \rightarrow \dbox$. Since we do not link $L_1$ and $L_2$, we should set $s=0$.
\begin{equation}\label{slidestoloop}
\begin{aligned}
\includegraphics[width=4.4in]{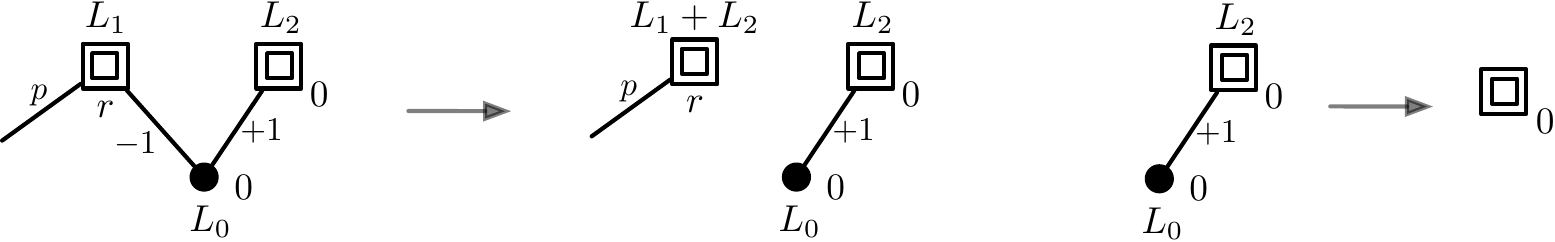}
    \end{aligned}
\end{equation}
The framing number for $\dbox$ means that if we slightly deform the matter circle $L_2$, there is no linking number between $L_2$ and its deformed circle. In other words, $L_2$ is an unknot without any twists.
Note that the definitions of framing numbers for matter circles $\sqbox$ or $\dbox$ and surgery circles are different. The latter is through the mapping class group, but the former is through the number of twists. In addition, on the neighborhood of the matter circles, one can always define an identical surgery which has framing number $\inf$.
The reversed movement of the above graphs is just \eqref{fusionchain}. A free hypermultiplet $\dbox$ often appear in the various identities, which is a subtle issue that we have not solved.


In (4.55) and (4.56) of \cite{witten89}, Witten provides formulas for the connect sum of Wilson loops, which we think could help us understand fusion identity, as the connected sum can be interpreted as a special handle slide as we show above. We copy Witten's formulas below:
\begin{align}
Z(S^3, \redLI \# \redLII ) \cdot Z(S^3, \redcirc)  &= Z(S^3, {\color{red}L_1})\cdot Z(S^3, {\color{red}L_2})  \label{firstformula}   \,, \\
Z(M_1+M_2, \redLI\#\redLII) \cdot Z(S^3, \redcirc) &= Z(M_1, \redLI)\cdot Z(M_2, \redLII) \label{secondformula}  \,.
\end{align}
Suppose that Wilson loops belong to links $\redcirc_1 \subset \redLI$ and $\redcirc_2 \subset \redLII$ respectively, and then \eqref{firstformula} mainly describes that the connected sum $\redLI\#\redLII$ is obtained through the connected sum of $\redcirc_1$ and $\redcirc_2$, which is the fusion of $\redcirc_1$ and $\redcirc_2$, and $\redcirc$ in $Z(S^3, \redcirc)$ is the connected sum $\redcirc_1 \# \redcirc_2$.  
The \eqref{secondformula} is the extension of the this connected sum of Wilson loops to generic three-manifolds. Witten's formula tells us that the connection sum of Wilson loops factorize.

Let us only analyze a simple example to see how the Verlinde coefficients join the story. 
Denoting $\redLI=\redcirc_1$ and $\redLII=\redcirc_2$, we can represent \eqref{slidestoloop} by the following formula:
\begin{align}\label{wittenslides}
Z(M_1, \redLI \# \redLII) \cdot Z(S^3, \bigcirc_0 \cup \redLII) = Z(M_1, \redLI \cup \bigcirc_0 \cup \redLII) \,,
\end{align}
where $Z(S^3, \bigcirc_0 \cup \redLII) = Z(S^2 \times S^1, \redLII)$, as the connect sum $S^3 \# \bigcirc_0 =\bigcirc_0 =S^2 \times S^1$. 
Suppose $\redLI$ has a representation $R_j$, $\redLII$ has a representation $R_k$, and $\bigcirc_0$ is associated with the representation $R_i$ before summing it up. It is defined in \cite{witten89} that $Z(S^2 \times S^1, R_a) =\delta_{a 0}$, $Z(S^2 \times S^1, R_a,R_b) =g_{ab} = \delta_{b,\bar{a}}$,  $Z(S^3) =S_{0,0}$ and $Z(S^3, L(R_i,R_j)) =S_{ij}$ where $L(R_i,R_j)$ denotes the Hopf link of $\redLI \cup \redLII$.
 We assume $M_1 = S^3_{+ 1}$. Then $Z(S^3, \bigcirc_0 \cup \redLII) =\sum_i S_0^iZ(S^2\times S^1, R_i, R_k) = \sum_i S_0^i g_{ik}=S_{0k} =Z(S^3, \redLII) $ which explains the duality $\bullet_0-\dbox = \dbox$. The right hand side $Z(M_1, \redLI \cup \bigcirc_0 \cup \redLII) = \sum_{i} S_{0}^iZ(S^3, L(R_i, R_j,R_k)) = \sum_{i} S_{0}^i  \sum_{m} S_{i}^m  Z(S^2\times S^1, R_m,R_j,R_k)  =\sum_{i,m}S_0^i S_i^m N_{mjk} $ where $L(R_i,R_j,R_k)$ denotes the Hopf link $\redLI \cup {\color{red} L_i}  \cup \redLII$. Therefore, the connected sum is represented as
\begin{align}
Z(M_1, \redLI \# \redLII)  =\frac{ \sum_{i,m}S_0^i S_i^m N_{mjk}  }{  S_{0k} } 
= \sum_i  \frac{S_0^i S_{ij}S_{ik}}{S_{0i}} \,,
\end{align}
where $N^{m}_{jk} =\sum_i \frac{ (S)^{-1}_{im} S_{ij}S_{ik}  }{ S_{0i}}$ and $\frac{ S_{ij} S_{ik}}{ S_{0i} } =\sum_m S_{i} ^{m} N_{m jk}$.

\subsection{Invariants of Hopf links}
If the linking number between two matter circles $L_1$ and $L_2$ are not zero, then we get a Hopf link $L_1 \cup L_2$. In e.g. \cite{Gukov:2016ac}, the colored HOMFLY invariants of the Hopf link is shown as 
\begin{align}\label{hopflinkP}
P_{[r_1],[r_2]} (T_{2,2}) = 
\frac{ (x,\q)_{r_1} }{ (\q,\q)_{r_1}} \cdot \sum_{i=1}^{r_2} (-\sqrt{\q})^{2 r_1 i} z^i  \cdot \frac{ (y, \q)_i}{(\q, \q)_i}  \,,
\end{align}
where $z=\q$ and $x= a $ and $y=a/\q$ and $a$ is the volume of the three-sphere.  The linking number is $L_1 \cdot L_2 =1$.
If combining $(-\sqrt{\q})^{2 r_1 i} z^i = ( \q^{r_1} z)^i$, one can see that this term comes from the gauging of the mass parameter $z \rightarrow z \q^{r_1}$. It would be interesting to compare this invariant with fusion identity \eqref{fusion}, but there are sill some barriers to directly relate this invariant to fusion identity.

The unrefined invariant is not complete. In the following, we will argue that the refinement is not a choice but compulsory. 
The refined invariant shows more interesting property, for which $x=\q t^{-2}$, $y=-a/\q t^3$. The refined HOMFLY reflect a more clear geometric interpretation that it is the coupling of two gauge circles:
\begin{equation}
\label{Hopfgraph}
\begin{aligned}
P^{\text{ref.}}_{[r_1],[r_2]} (T_{2,2}) &= \frac{ (xy,\q)_{r_1}}{(x,\q)_{r_1}} \cdot \sum_{i=0}^{r_2} (-\sqrt{\q})^{2 r_1 i} x^i \cdot \frac{ (y,\q)_{i}}{(x \q^{r_2-i},\q )_i} \begin{bmatrix} r_2\\i  \end{bmatrix}_{\q} \\
&=  \frac{ (xy,\q)_{r_1}}{(x,\q)_{r_1}} \cdot \sum_{i=0}^{r_2} 
(-\sqrt{\q})^{2 r_1 i} x^i \cdot
\frac{ (x,\q)_{r_2-i}}{(\q,\q)_{r_2-i}} \frac{(y,\q)_i}{(\q,\q)_i}   \,.
\end{aligned}
\end{equation}
Graphically, 
\begin{equation}
\begin{aligned}
\begin{tikzpicture}
\draw[thick] (0,0)--(-0.5,1);
\node at (0,0) {\gauge}; 
\node at (0,-0.4) {$r_1$};
\draw[thick] (-0.5,1)--(-1,0)--(-1.5,1)  ;
\node at (-0.5,1){\dbox};
\node at (-1.5,1) {\dbox};
\node at (-1,0) {\gauge} ;
\node at (-1,-0.4) {$j$};
\draw[thick] (0,0)--(2,0)node [midway, above]{\small $+1$}node{\gauge}--(1.5,1) node {\dbox} ; \node at (2,-0.4)  {$i$};
\draw[thick] (3,0)node {\gauge}--(2.5,1)node {$\dbox$}--(2,0)  ;
\node at (3,-0.4){$r_2$};
\end{tikzpicture}
\end{aligned}
\end{equation}
Because of this, the refinement parameter $t$ should be viewed as a mass parameter in our context, rather than a parallel parameter as $\q$.
When $x=\q$, the above reduces to the unrefined form in \eqref{hopflinkP}. We have not clearly understood this zig-zag move for the Hopf link through geometry. We can only observe that each cross of Hopf links may corresponding to a fusion. Note that the linking between $\bullet_{r_1}-\bullet_{i}$ can be switched to $\bullet_j-\bullet_{r_2}$ as this Hopf link is symmetric.
For the Hopf link with three components $L_1\cup L_2 \cup L_3$, one can just extend the graph in \eqref{Hopfgraph} by adding one more repeated component.

One can also read off some features for a knot, such as the HOMFLY-PT of a trefoil is 
\begin{align}
 \sum_{r=0}^\inf \frac{\tx^r}{(\q,\q)_r} \cdot P_{[r]}(3_1) = \sum_{r=0}^\inf \tx^r \cdot \sum_{k=0}^r  (-\sqrt{\q})^{2 k r}  x^k \cdot \frac{ (z,\q)_{r-k} }{ (\q,\q)_{r-k} } \cdot \frac{(y,\q)_k }{ (\q,\q)_k}  \,,
\end{align}
where $y=-a t/\q$, $x =\q t^2$ and $z=0$. 

These examples indicate the invariants of knots are diverse couplings of hypermultiplets.

\section{Open questions}\label{sec:open}

There are still many open questions:
\begin{enumerate}
\item The fusion of matter circles is the key to other dualities, and indicates that fusion may also happen when two components of a knot approach each other. This shows a similarity between fusion (connected sum) and R-matrix. Then the question is what is the relation between fusion and R-matrix, or equivalently is there a more generic fusion that describes the crosses in knots.

\item What is the algebra that describes the duality structures of 3d theories determined by the three-manifolds? Is it a modified Hopf algebra?

\item How to describe abelian theories with complicated superpotentials? Is it possible to use the hiding algebraic structures to describe them systematically?

\item Does the fusion identity explain non-abelian theories?

\item Why does the  FI parameter of $\bullet_0$ equal to a mass parameter $x$ in the fusion identity \eqref{fusion}? Does this mean superpotentials?

\item  Could theta functions and superconformal indices be well defined for the handle slides? 

\item How to understand Verlinder formula and the level-rank duality  in terms of three-manifolds and Ooguri-Vafa defects?

\end{enumerate}

\acknowledgments    
I thank Sergei Gukov, Chiung Hwang, Piotr Kucharski, Sung-Soo Kim, Ioannis Lavdas, Kimyeong Lee, Satoshi Nawata, Pichai Ramadevi, and Piotr Su\l{}kowski for helpful discussions and correspondences. I also thank YITP for “The 18-th Kavli Asian Winter School (YITP-W-23
-13)”, Banach Center for “The summary workshop on Knots, homologies and physics”, and Shanghai University for “2024 Strings, gravity and gravitational waves”, and UESTC and SCU for the workshop “Tianfu Fields and Strings 2024”, where the author was partly supported. 
This work is supported by NSFC Grant No.11850410428 and NSFC Grant No.12305078.

\newpage

\appendix
\section{$\q$-Pochhammer products}
We show some useful identities below
\begin{align}
& (x, \q)_n = \prod_{k=0}^{n-1} (1-x \q^k)   \,,  \\
& (x \, \q^m , \q)_n = \frac{(x, \q)_{m+n}}{ (x, \q)_m  }    \,,\\
&  (x \, \q^m , \q)_n = \frac{(x  \, \q^m, \q)_{\infty
}}{ (x \, \q^{ m +n }, \q)_\infty  }   \,, \\
& \frac{1}{ ( x, \q )_n } = \frac{ (x \, \q ^n , \q)_\infty  }{ (x , \q)_\infty }   \,, \\
& (x, \q)_n =\frac{ (x, \q)_\inf}{(x \q^n,\q)_\inf }    \,, \\
& (x \q^n,\q)_\inf= \frac{ (x,\q)_\inf }{ (x,\q)_n } =  \sum_{d=0}^{\inf} (-\sqrt{\q})^{2 n d +d^2} \frac{(x/\sqrt{\q})^d}{ (\q,\q)_d}  \,,   \\
&  (x,\q)_\inf=  \PE \left[  - \frac{x}{1-\q} \right] \,,~~~~
\frac{1}{ (x,\q)_\inf }  = \PE \left[   \frac{x}{1-\q} \right]   \,,
  \\
  &   (x,\q)_n = \PE \left[   \frac{ x \q^n -x  }{1-\q}  \right] \,,~~~~
  \frac{ (x,\q)_\inf}{   (x,\q)_n } =\PE \left[   -\frac{x \q^n}{1-\q} \right]   \,.
\end{align} 
Because of the property $\PE[ f (x)] \PE[f (y)] = \PE[f(x+y)]$, we can get
\begin{align}
(x,\q)_n (y,\q)_n = (x+y,\q)_n \,,~~ \frac{(x,\q)_n}{ (y,\q)_n} = (x-y,\q)_n =\frac{1}{ (y-x,\q)_n }   \,.
\end{align}
Some notations in quantum groups are
\begin{align}
&   [n]_\q =\frac{ 1- \q^n}{1-\q } \,,~~~~  [n]_\q ! = \prod_{k=1}^n [k]_\q= \frac{(\q,\q)_n}{ (1-\q)^n}     \,,\\
& 
 \bigg[\begin{matrix}   n\\k \end{matrix} \bigg]_\q  := \frac{ [n]_\q !  }{ [n-k]_\q !  [k]_\q !  } \,,~~~~
  \lim_{q \rightarrow 1} [n]_\q! =n! \,,~~ \lim_{\q \rightarrow 1} [n]_\q =n   \,.
\end{align}
We can rewrite using MacMahon functions $
(\q x,\q)_\inf = {M(x)}/{M(\q x)}  \,,
$
which leads to 
\begin{align}
\frac{(\q x,\q)_\inf}{(x,\q)_\inf} = \frac{M(x/\q)}{M(\q x)} \,.
\end{align}

\subsection*{Bailey's lemma.}
The Bailey's lemma is the relation between a pair $(\a_n,\b_n)$ by 
\begin{align}
\b_n = \sum_{k=0}^n \frac{\a_k}{ (\q,\q)_{n-k} (\q x,\q)_{n+k}  }  \,.
\end{align}
Equivalently, it is
\begin{align}
\a_n \cdot \frac{(x,\q)_n }{ (\q x,\q )_n }  = \sum_{k=0}^n \frac{(x,\q)_k }{(\q,\q)_k} \cdot (-\sqrt{\q})^{k^2} (1/\sqrt{\q})^k \cdot \b_{n-k}    \,.
\end{align}
The duality \eqref{sdualgauge} is a Bailey pair when $z=1$,
\begin{align}
\a_n =\frac{ (1, \q)_\inf}{ (x,\q)_\inf}  \,,~~\b_{n-k} = \frac{1}{ (\q x,\q)_{n-k}  (x,\q)_k}  \,,
\end{align}
in which although $ (1,\q)_\inf$ looks a problem.

\newpage

\bibliographystyle{JHEP}
\bibliography{refgeoplumb}

\end{document}